\documentclass{ieeeaccess}

\usepackage{cite}
\usepackage{amsmath,amssymb,amsfonts}
\usepackage{algorithmic}
\usepackage{graphicx}
\usepackage{textcomp}

\usepackage[hyphens]{url}

\usepackage{pifont}

\usepackage{siunitx}
\usepackage{microtype}

\usepackage{hyperref}

\usepackage{booktabs} 	

\usepackage{colortbl}   

\usepackage{scrlayer} 

\definecolor{platinum}{rgb}{0.9, 0.89, 0.89}
\definecolor{alizarin}{rgb}{0.82, 0.1, 0.26}

\newcommand{\orcid}[1]{\href{https://orcid.org/#1}{\includegraphics[width=8pt]{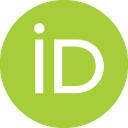}}}

\DeclareMathOperator*{\argmin}{arg\,min} 

\def\BibTeX{{\rm B\kern-.05em{\sc i\kern-.025em b}\kern-.08em
    T\kern-.1667em\lower.7ex\hbox{E}\kern-.125emX}}

\usepackage{draftwatermark}
\SetWatermarkText{This is the \\ prepublication \\ version}
\SetWatermarkScale{0.6}

\begin{document}
\history{Received November 14, 2021, accepted December 20, 2021, date of current version January 3, 2022.}
\doi{10.1109/ACCESS.2022.3141198}

\title{Toward Generating Sufficiently Valid Test Case Results: A Method for Systematically Assigning Test Cases to Test Bench Configurations in a Scenario-Based Test Approach for Automated Vehicles}

\author{\uppercase{Markus~Steimle}\,\orcid{0000-0001-8913-6980}\authorrefmark{1}, \uppercase{Nico Weber}\,\orcid{0000-0002-6753-6104}\authorrefmark{2}, \uppercase{and Markus~Maurer}\,\orcid{0000-0002-5357-9701}\authorrefmark{1}}
\address[1]{Institute of Control Engineering, Technische Universität Braunschweig, 
	38106 Braunschweig, Germany (e-mail: \{steimle, maurer\}@ifr.ing.tu-bs.de)}
\address[2]{Opel Automobile, Stellantis, 65423 Rüsselsheim am Main, Germany (e-mail: nico.weber@external.stellantis.com)}
\tfootnote{This work was supported by the German Federal Ministry for Economic Affairs and Energy within the project ``\textbf{S}imulationsbasiertes \textbf{E}ntwickeln und \textbf{T}esten von automatisiertem Fahren (SET Level)–Simulation-Based Development and Testing of Automated Driving,'' a Successor Project to the project ``\textbf{P}rojekt zur \textbf{E}tablierung von \textbf{g}enerell \textbf{a}kzeptierten Gütekriterien, Werkzeugen und Methoden sowie \textbf{S}zenarien \textbf{u}nd \textbf{S}ituationen zur Freigabe hochautomatisierter Fahrfunktionen (PEGASUS)'' and a project in the PEGASUS family.
}

\markboth
{Steimle~\emph{et~al.}: Toward Generating Sufficiently Valid Test Case Results}
{Steimle~\emph{et~al.}: Toward Generating Sufficiently Valid Test Case Results}

\corresp{Corresponding author: Markus Steimle (e-mail: steimle@ifr.ing.tu-bs.de).}

\begin{abstract}

To successfully launch automated vehicles into the consumer market, there must be credible proof that the vehicles will operate safely. 
However, finding a method to validate the vehicles' safe operation is a challenging problem.
While scenario-based test approaches seem to be possible solutions, they require execution of a large number of test cases.
Several test benches, ranging from actual test vehicles to partly or fully simulated environments, are available to execute these test cases.
Each test bench provides different elements, which in turn, have different parameters and parameter ranges. 
The composition of elements with their specific parameter values at a specific test bench that is used to execute a test case is referred to as a test bench configuration.
However, selecting the most suitable test bench configuration is difficult.
The selected test bench configuration determines whether the execution of a specific test case provides sufficiently valid test case results with respect to the intended purpose, for example, validating a vehicle's safe operation.
The effective and efficient execution of a large number of test cases requires a method for systematically assigning test cases to the most suitable test bench configuration.
Based on a proposed method for classifying test bench configurations, we propose and illustrate a method for systematically assigning test cases to test bench configurations in a scenario-based test approach for automated vehicles. 
This assignment method allows for the effective and efficient execution of a large number of test cases while generating sufficiently valid test case results.
\end{abstract}

\begin{keywords}
	Automated vehicles, 
	classification,
	PEGASUS family,
	scenario-based test approach,
	sufficiently valid test case results, 
	test bench configuration, 
	test case assignment,
	X-in-the-loop,
	test method.
\end{keywords}

\titlepgskip=-15pt

\maketitle

\section{Introduction} \label{sec_introduction}

\IEEEPARstart{V}{alidating} that a vehicle operates safely is a challenge when introducing automated vehicles with SAE levels~$\geq$~3~\cite{SAE_J3016_2021}~\cite{Riedmaier_2020}. 
Existing safety validation approaches, such as the statistical and distance-based safety validation approach that is currently applied to vehicles equipped with SAE level~$<$~3~\cite{SAE_J3016_2021} driving automation systems, are inapplicable due to time and cost constraints. 
Wachenfeld and Winner~\cite{Wachenfeld_2016b} call the application of existing safety validation approaches ``an `approval-trap' for autonomous driving.''
For this reason, new approaches for ensuring and validating the safe operation of automated vehicles are currently under development.
For example, there are scenario-based approaches, function-based approaches, formal verification approaches, shadow mode approaches, and the staged introduction of automated vehicles (e.g.,~\cite{Riedmaier_2020} and \cite{Stellet_2020}).
Scenario-based development and test approaches might be possible solutions to ensure and validate the safe operation of automated vehicles (e.g.,~\cite{Riedmaier_2020}, \cite{ISODIS21448_2021}, \cite{Schuldt_2017}, and \cite{Wood_2019}). 
These approaches have been investigated in recently completed research projects, for example, aFAS~\cite{Stolte_2015}, \mbox{ENABLE-S3}~\cite{ENABLE_S3_Method_2019}, and PEGASUS~\cite{PEGASUSMethod_2019}.
These scenario-based approaches are currently being pursued and investigated in ongoing research projects, for example, SET Level~\cite{SL45_Homepage} and VVM~\cite{VVM_Homepage}, which are successors to the PEGASUS project and projects of the PEGASUS family.

Menzel~\emph{et~al.}~\cite{Menzel_2018a} described how scenarios can be systematically evolved and used throughout the development process phases outlined in the ISO~26262 standard~\cite{ISO26262_2018}. 
The required scenarios can be derived using a data-driven approach, a knowledge-driven approach, or a combination of both~\cite{Menzel_2019}.
Data-driven approaches were presented, for example, by Gelder and Paardekooper~\cite{Gelder_2017}, Pütz~\emph{et~al.}~\cite{Putz_2017}, Waymo~LLC~\cite{waymo_SafetyReport_2020}, and Weber~\emph{et~al.}~\cite{Weber_2020}.
Bagschik~\emph{et~al.}~\cite{Bagschik_2018b} proposed a knowledge-driven approach for generating traffic scenes in natural language using an ontology as a basis for generating scenarios.
Bagschik~\emph{et~al.}~\cite{Bagschik_2018} extended this approach to implement a knowledge-based generation of operating scenarios for German highways.
Based on this work, Menzel~\emph{et~al.}~\cite{Menzel_2019} proposed an approach that automatically provides scenario descriptions to execute the generated scenarios in a simulation environment. 
Subsequently, these so-called concrete scenarios can be used in further steps to derive test cases.
However, how to derive the ``right'' test cases from the generated scenarios is still an open research question, which is addressed, for example, in the VVM project~\cite{VVM_Homepage}.
Neurohr~\emph{et~al.}~\cite{Neurohr_2020} presented an abstract framework around such a scenario-based test approach that includes the process step for deriving test cases.
Steimle~\emph{et~al.}~\cite{Steimle_2021} proposed a structuring framework and a basic vocabulary for scenario-based development and test approaches for automated vehicles. 
This structuring framework is based on a generic development and test process that also includes the process step of deriving test cases.
Other than the process step, Neurohr~\emph{et~al.}~\cite{Neurohr_2020} and Steimle~\emph{et~al.}~\cite{Steimle_2021} have not precisely described how these test cases will be derived.
Baumann~\emph{et~al.}~\cite{Baumann_2021} presented an approach to generate critical test cases based on scenarios using a reinforcement learning algorithm to reduce the total number of test cases to be simulated.

Applying the methods mentioned above leads to a large number of test cases.
However, it is still unclear how many test cases will be derived.
For example, considering a self-defined number of parameters and self-defined instances per parameter for a functional scenario, Amersbach~\cite{Amersbach_2020} calculated $10^{31}$ possible test cases for the exemplary scenario set he considered.
Applying the method he proposed to reduce this number would still result in $1.2 \cdot 10^{8}$ test cases.
Baumann~\emph{et~al.}~\cite{Baumann_2021} calculated $1.8 \cdot 10^{37}$ test cases by pure combinatorics of the parameters in the overtaking assistant they considered. 
This number of test cases was reduced using a reinforcement learning algorithm.
However, they did not specify the exact number of test cases that must finally be executed to validate the safe operation of the vehicle.
Although the numbers mentioned above should be taken with caution, as many assumptions are included in their calculation, they show that a large number of test cases need to be executed in a sufficiently valid manner\footnote{In this publication, we follow the recommendation of Balci~\cite{Balci_2010} to use the term ``sufficiently valid'' instead of ``(absolutely) valid'' because, according to Law~\cite{Law_2019}, there is no such thing as absolute validity for a simulation model, nor is it even desired. 
``Sufficiently valid'' indicates that the validity has been judged with respect to the intended purpose, for example, validating a vehicle's safe operation, and has been found to be sufficient.}.

Several test benches, ranging from actual test vehicles to partly or fully simulated environments, are available to execute the derived test cases.
However, selecting the most suitable test bench and its configuration, for example, in terms of the required execution time and costs, is challenging.
There are several reasons for using simulative test methods instead of real-world driving tests if they are applicable and sufficiently valid.
These reasons include, for example, that test cases can be executed before prototypes or target hardware are available, that test case execution can be faster or slower than real time, that the simulation models developed can be used several times, that the risk to people and material is significantly reduced when testing safety-relevant driving functions, and that it might be possible to reduce costs in the development process.
Therefore, it seems that test cases should be executed using simulative test methods whenever possible.
However, to be able to make a reliable assertion with the generated test case results, for example, in regard to a vehicle's safe operation, which is a prerequisite for its market release, it is crucial that the generated test case results are sufficiently valid.

Whether sufficiently valid test case results can be generated with a specific test bench depends on the test bench's characteristics.
Every test bench provides various elements that need to be interconnected to execute test cases. 
For example, assume a software-in-the-loop test bench that provides, in addition to other simulation models, two vehicle dynamics simulation models (e.g., a simple single-track simulation model and a more complex double-track simulation model).
Each of these vehicle dynamics simulation models has a different accuracy and execution time and they can be interchanged depending on the test case.
Consequently, the simple single-track simulation model may be sufficiently valid for one test case, while another test case may require the more complex double-track simulation model.
The composition of all the elements with their specific parameter values at a specific test bench that is used to execute a test case is referred to as a test bench configuration. 
This composition determines the actual performance\footnote{In this context, the performance of a test bench (configuration) is understood as the sum of the characteristics of the test bench (configuration) that influences the (effective and efficient) execution of test cases. 
Based on Walden~\emph{et~al.}~\cite{Walden_2015}, these characteristics can be physical and/or functional attributes.
According to Walden~\emph{et~al.}~\cite{Walden_2015}, these ``performance attributes include quantity (how many or how much), quality (how well), timeliness (how responsive, how frequent), and readiness (when, under which circumstances).''
Thus, performance attributes in this context include, for example, the ability to generate sufficiently valid test case results -- including the accuracy of the test case results -- and the time needed to generate these test case results.} of this test bench configuration.
Therefore, this composition determines whether the execution of a specific test case with this test bench configuration provides sufficiently valid test case results.
When using simulative methods for test case executions, the simulation models used do not capture all the attributes of the real counterpart they represent. 
They capture only those attributes that seem relevant to the respective simulation model's developers and/or simulation model's users for the intended purpose.
The sum of those attributes that seem relevant for the intended purpose is referred to as the \textit{reality of interest} of the real counterpart in the NASA~7009A standard~\cite{NASA_7009A_2016}.
To maintain their efficiency, simulation models should not be more detailed than necessary for their intended purpose~\cite{Klemmer_2011}.
Thus, the simulation models used to execute a test case directly affect the actual performance of the test bench configuration to which they belong.

In statistical and distance-based test approaches, as applied to vehicles equipped with SAE level~$<$~3~\cite{SAE_J3016_2021} driving automation systems, test cases are usually executed in real-world driving tests with various test vehicles~\cite{Wachenfeld_2016b}.
When simulative test approaches are used for test case execution, experts usually manually decide which simulation models and thus which test bench configuration should be used. 
Therefore, experts decide which test bench configuration is sufficiently valid.
To check this validity after executing a test case, it is still common practice for experts to manually compare graphs of simulation data and graphs of measurement data generated during driving tests~\cite{Viehof_2018b}.
This practice has two challenges.
On the one hand, the expressiveness of this manual and subjective method is limited in its ability to validate an automated vehicle's safe operation. 
This limitation means that there is usually no objective evidence that the test case results are sufficiently valid for a safety validation. 
On the other hand, because of the expected large number of test cases and the various elements provided at test benches, it is ineffective and inefficient to manually select test bench configurations for each test case. 
Therefore, an effective and efficient execution of a large number of test cases requires a method for systematically (and at best automatically) assigning test cases to the most suitable test bench configuration.
Additionally, it must be considered whether this test bench configuration is sufficiently valid.

There are methods for assigning test cases to simulative test methods (Schuldt~\emph{et~al.}~\cite{Schuldt_2015} and Schuldt~\cite{Schuldt_2017}), for splitting the execution of test cases into simulation-based and reality-based tests (Böde~\emph{et~al.}~\cite{Bode_2018}), and for making recommendations when choosing among four test methods (Antkiewicz~\emph{et~al.}~\cite{Antkiewicz_2020}).
However, assigning test cases to test methods is not expedient because a test method alone does not provide enough information about the actual performance of the specific test bench configuration used to execute a test case (see Section~\ref{ssec_related_work_assignment}).
Nevertheless, the methods mentioned above can help to exclude test methods in advance based on the generic characteristics of a test method. 
Accordingly, these test methods cannot be used to execute the respective test cases.
However, for the systematic assignment of test cases to test bench configurations, the elements provided at a specific test bench must be considered, which is impossible with the methods mentioned above.
Only by considering the elements provided at a test bench and thus the actual performance of the test bench configuration resulting from the composition of these elements, is it possible to generate sufficiently valid test case results effectively and efficiently.
To date, we are not aware of any publication that proposes a method suitable for such an assignment.

For the systematic assignment of test cases to test bench configurations, a method for classifying test benches and test bench configurations is indispensable in formalizing the characteristics of a test bench and, in particular, the characteristics of its provided test bench configurations.
There are generic approaches for classifying test methods (Schuldt~\emph{et~al.}~\cite{Schuldt_2015}, Schuldt~\cite{Schuldt_2017}, and Antkiewicz~\emph{et~al.}~\cite{Antkiewicz_2020}) and test benches (Strasser~\cite{Strasser_2012}, von Neumann-Cosel~\cite{NeumannCosel_2014}, and Stellet~\emph{et~al.}~\cite{Stellet_2015}). 
However, these generic classification methods do not consider the specific elements provided at a specific test bench (Antkiewicz~\emph{et~al.}~\cite{Antkiewicz_2020} uses a slightly consolidated list of test elements), for example, various simulation models (see Section~\ref{ssec_related_work_classification}). 
A method for classifying test benches and test bench configurations that can be used as a basis to assign test cases to test bench configurations systematically, must necessarily consider the elements provided at a specific test bench. 
This consideration is not possible with the methods mentioned above.
Thus far, we are not aware of any publication that proposes a method for classifying test bench configurations that can be used as a basis to assign test cases to test bench configurations systematically.

In this publication, we propose a method for classifying test benches and test bench configurations based on a publication we have previously published on arXiv~\cite{Steimle_2019}.
As a basis for this classification method, we systematically derive the functionalities that a test bench must provide to execute a test case.
These functionalities are identified by analyzing the functional system architecture of an automated vehicle and its interactions with the driver or user and the environment during the execution of a test case.
Based on this classification method, we propose a method for systematically assigning test cases to test bench configurations, which allows for the effective and efficient execution of test cases while generating sufficiently valid test case results.
This method forms the basis for an implementation that will allow for the effective, efficient, and automated execution of extensive test case catalogs.
For discussion purposes, an earlier version of this publication was published on arXiv in September 2021~\cite{Steimle_2021a}.
We also solicited feedback on our approaches from research and industry partners in the projects SET Level~\cite{SL45_Homepage} and VVM~\cite{VVM_Homepage}.
This feedback is discussed and considered in this version.

\subsection*{Novelty and main contributions to the state of the art} 

The novelty and the main contributions of this publication are not only a method for classifying test benches and test bench configurations but also a method for systematically assigning test cases to test bench configurations based on the classification method.
Therefore, we

\begin{itemize}
    \item identify requirements that the methods should fulfill.
    \item review relevant literature on various classification and assignment methods.
    \item evaluate their suitability for classifying test benches and test bench configurations and their suitability for systematically assigning test cases to test bench configurations.
    \item propose and describe a method for classifying test benches and test bench configurations.
    \item propose and describe a method for systematically assigning test cases to test bench configurations.
    \item explain the application of the two proposed methods by means of examples.
    \item highlight and discuss some of the current limitations and challenges of using simulation models and emulated elements for test case execution.
    \item identify further research that is needed to execute extensive test case catalogs effectively and efficiently in an automated manner while generating sufficiently valid test case results.
    \item evaluate the proposed methods based on the requirements identified. 
\end{itemize}

\subsection*{Structure}

This publication is structured as follows:
Section~\ref{sec_requirements} presents the requirements that the method for classifying test benches and test bench configurations shall fulfill. 
Additionally, it presents the requirements that the method for systematically assigning test cases to test bench configurations should fulfill.
Section~\ref{sec_related_work} provides a brief motivation for the methods we propose. 
This motivation is based on selected related work regarding methods for classifying test methods and test benches and selected related work regarding methods for assigning test cases. 
Section~\ref{sec_classification_method} describes the proposed method for classifying test benches and test bench configurations, which is evaluated in Section~\ref{sec_examples_classification} by illustrating various examples of classified test benches and test bench configurations.
Section~\ref{sec_assignment_method} describes the proposed method for systematically assigning test cases to test bench configurations, which is evaluated in Section~\ref{sec_example_assignment} by assigning a test case to a test bench configuration according to the proposed assignment method.
Section~\ref{sec_req_evaluation} evaluates the methods we propose based on the requirements identified.
Finally, Section~\ref{sec_conclusion} presents our conclusions and describes future work.

\section{Requirements Specification} \label{sec_requirements}

In this section, we present the requirements that the method for classifying test benches and test bench configurations shall fulfill (denoted with a C).
Additionally, we present the requirements that the method for systematically assigning test cases to test bench configurations shall fulfill (denoted with an A).

\subsection{Classification Method} \label{ssec_req_classificationMethod}

We identified the following requirements for the classification method:

\begin{itemize}
    \item The classification method shall allow systematic~(Req.~C1a) and objective~(Req.~C1b) classification of test benches and test bench configurations. 
    
    \item The classification method shall consider the elements provided at a particular test bench~(Req.~C2).
    
    \item The criteria used for the classification shall be derived systematically~(Req.~C3).
    
    \item The classification method shall be as simple and easily understandable as possible~(Req.~C4). 
    
    \item The classification method shall be automatable~(Req.~C5). 
    
    \item The classification method shall allow the generation of machine-readable classification results~(Req.~C6).
    
    \item The classification method shall allow intuitive visualization of the classification results~(Req.~C7).
    
\end{itemize}

\subsection{Test Case Assignment Method} \label{ssec_req_assignmentMethod}

We identified the following requirements for the test case assignment method:

\begin{itemize}
    \item The test case assignment method shall allow systematic~(Req.~A1a) and objective~(Req.~A1b) assignment of test cases to test bench configurations.
   
    \item The test case assignment method shall allow the effective~(Req.~A2a) and efficient~(Req.~A2b) generation of sufficiently valid test case results.
    
    \item The test case assignment method shall be as simple and easily understandable as possible~(Req.~A3).
 
    \item The assignment method shall be automatable~(Req.~A4). 
    
    \item The test case assignment method shall already be applicable with expert knowledge~(Req.~A5) (see note~1).
    
\end{itemize}

\textit{Note~1 (belonging to Req.~A5)}:
This requirement is especially relevant to this publication.
As we will see later, there is still a need for research on some input artifacts of the proposed test case assignment method for full automation of this method. 
This research is not part of this publication. 
Thus, the method must be supported by expert knowledge at the moment.
Therefore, a particular requirement within this publication is that the test case assignment method shall already be applicable with expert knowledge.%

\section{Related Work} \label{sec_related_work}

\subsection{Classification Methods} \label{ssec_related_work_classification}

Existing studies have classified test benches and test methods based on various characteristics but the characteristics used for these classifications vary from author to author.
In this section, we describe and evaluate five classification methods as examples. 
Some of the publications are only available in German language. 
Therefore, we describe the methods proposed in these publications in more detail.
Additionally, we list the test methods or test benches classified by the respective authors.
In this publication, we understand a test bench to be the ``technical device,'' which consists of software and hardware that can be used to execute test cases.
For example, using a specific software-in-the-loop test bench, a test case can be executed while software (i.e.,~the test object) is tested in the loop.
Furthermore, we understand a test method to be the way in which a test case can be executed without referring to a specific test bench. 
For example, if a test case is to be executed while software (i.e.,~the test object) is tested in the loop (independent of a specific test bench), the corresponding test method is called software-in-the-loop.
Therefore, a test bench implements a test method.

Strasser~\cite{Strasser_2012} classified test benches based on the structure of a human-machine system, as shown in Fig.~\ref{fig_Struktur_Strasser}. 
This structure consists of four modules: the environment, driver, vehicle, and electric vehicle system that represents the system under test. 
Each module can be simulated or can be real.
By combining simulated and real modules, Strasser~\cite{Strasser_2012} classified test benches as software-in-the-loop, hardware-in-the-loop, driver-in-the-loop, vehicle-in-the-loop, and test vehicles (onboard test or rapid prototyping).
With this binary classification of simulated and real modules, however, it is not possible to uniquely classify every possible element of a test bench. 
As an example, Schuldt~\emph{et~al.}~\cite{Schuldt_2015} mentioned a balloon vehicle that is neither a real nor a simulated vehicle but is emulated by hardware with similar dimensions. 
This distinction may be significant for testing sensors, as the sensor data representing the balloon vehicle may differ from the sensor data of a real or a simulated vehicle.
For this reason, based on Wachenfeld and Winner~\cite{Wachenfeld_2016b}, Schuldt~\emph{et~al.}~\cite{Schuldt_2015} proposed classifying such elements~--~elements that are neither the real element nor the simulated counterpart~--~as emulated elements.

Stellet~\emph{et~al.}~\cite{Stellet_2015} classified test methods by dividing the overall complex of vehicle, environment, and driver blockwise (sequentially) into real-world and virtual data (i.e.,~simulation).
Different test methods can be classified by color-coding the respective blocks (according to whether they are part of the real world or the simulation).
Fig.~\ref{fig_Struktur_Stellet} shows a neutral representation without color-coded blocks.
It was unclear to us why Stellet~\emph{et~al.}~\cite{Stellet_2015} did not draw an arrow between the driver and the vehicle and arrows between the blocks inside the vehicle. 
After consulting with Stellet~\cite{Stellet_2021}, we found that these arrows should actually be present.
Based on the division mentioned above, Stellet~\emph{et~al.}~\cite{Stellet_2015} classified test methods as hardware-in-the-loop, driver-in-the-loop, vehicle-hardware-in-the-loop, and vehicle-in-the-loop.
Stellet~\emph{et~al.}~\cite{Stellet_2015} also did not consider emulated components.

Von Neumann-Cosel~\cite{NeumannCosel_2014} classified test benches based on the interaction of a driving function (a combination of sensors, algorithms, and actuators) with the environment, driver, and vehicle, as shown in Fig.~\ref{fig_Struktur_Neumann}.
It is noteworthy that it is unclear to us whether the ``restraint devices'' belong to the arrow pointing to the vehicle or the driver, because, as depicted in Fig.~\ref{fig_Struktur_Neumann}, the ``restraint devices'' are located exactly between these two arrows. 
Von Neumann-Cosel~\cite{NeumannCosel_2014} did not describe the relationship of the ``restraint devices.''
Each of the components mentioned above can be simulated or real.
Von Neumann-Cosel~\cite{NeumannCosel_2014} also did not consider emulated components.
Based on the combination of simulated and real components, von Neumann-Cosel~\cite{NeumannCosel_2014} classified test benches as concept-in-the-loop, software-in-the-loop, hardware-in-the-loop, driver-in-the-loop, vehicle-in-the-loop, and a test vehicle equipped with a driving robot that operates the steering wheel and pedals. 
Thus, this driving robot emulates the driver behavior.

\Figure[t!](topskip=0pt, botskip=0pt, midskip=0pt)[scale=0.98]{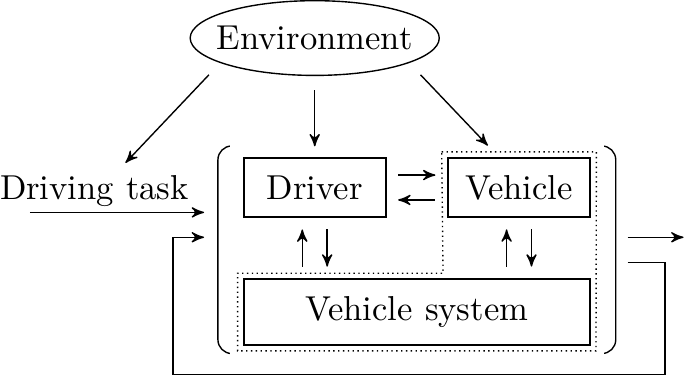}
{Structure of a human-machine system based on Strasser~\cite{Strasser_2012} based on Bubb and Schmidtke~\cite{Bubb_1993}.\label{fig_Struktur_Strasser}}

\Figure[t!](topskip=0pt, botskip=0pt, midskip=0pt)[]{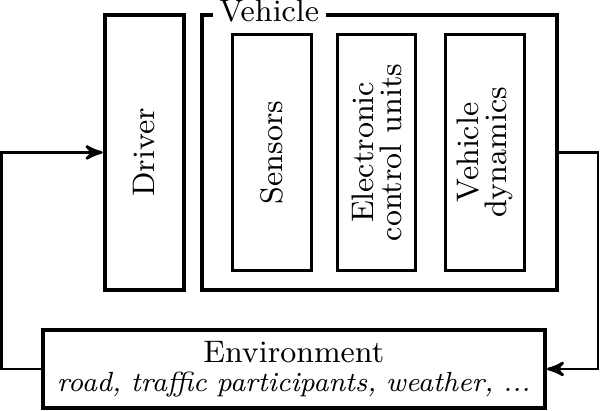}
{The overall complexity of the vehicle, environment, and driver (each block can be part of the simulation or the real world) based on Stellet~\emph{et~al.}~\cite{Stellet_2015}; different test methods can be classified by color-coding the blocks according to the test method under consideration (in this figure: neutral representation without color-coded blocks).\label{fig_Struktur_Stellet}}

\Figure[t!](topskip=0pt, botskip=0pt, midskip=0pt)[scale=0.98]{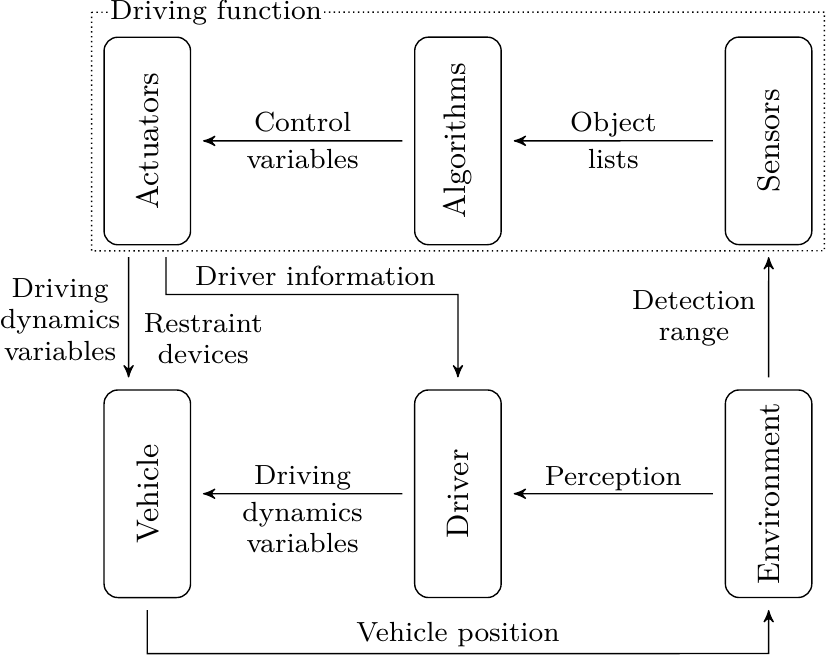}
{Interaction of the driving function with the environment, driver, and vehicle based on von Neumann-Cosel~\cite{NeumannCosel_2014}.\label{fig_Struktur_Neumann}}

Based on Schuldt~\emph{et~al.}~\cite{Schuldt_2015}, Schuldt~\cite{Schuldt_2017} classified test methods according to different dimensions: the test object, driver behavior, (residual) vehicle, vehicle dynamics, (residual) perception, road users, and scenery.
Each dimension can be simulated, emulated, or real.
By combining these dimensions, Schuldt~\cite{Schuldt_2017} classified test methods as software-in-the-loop, hardware-in-the-loop, driver-in-the-loop, vehicle-hardware-in-the-loop, and vehicle-in-the-loop.
To visualize the test methods, Schuldt~\cite{Schuldt_2017} used radar charts in which each spoke represents a dimension of the test method.
Fig.~\ref{fig_KV_empty} shows a blank radar chart.

\Figure[t!](topskip=0pt, botskip=0pt, midskip=0pt)[]{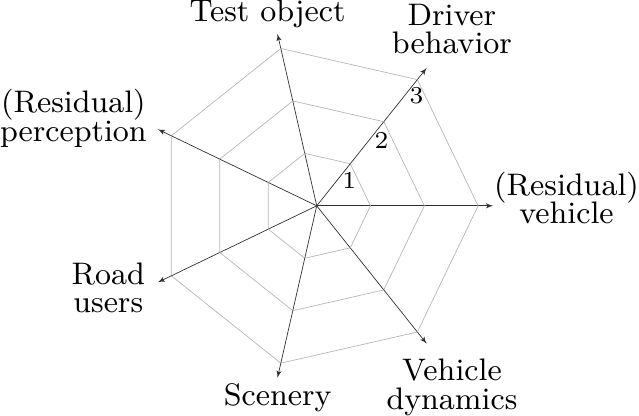}
{Blank radar chart with stages 1~=~simulated, 2~=~emulated, and 3~=~real based on Schuldt~\cite{Schuldt_2017}.\label{fig_KV_empty}}

Antkiewicz~\emph{et~al.}~\cite{Antkiewicz_2020} investigated and compared four modes of scenario testing: closed-course testing with real actors (e.g., vehicles and pedestrians), closed-course testing with robotic surrogate (i.e., dummy) actors, closed-course testing with mixed reality, and simulation testing.
Within the paper, they do not consider open-road testing, ``as it differs sufficiently from scenario testing and it requires a different testing method.''
The term \textit{surrogate} corresponds to what we mean by the term \textit{emulated}, and the \textit{mode of (scenario) testing} is what we mean by a \textit{test method (for scenario-based testing)}.
These four modes of scenario testing are classified based on real and simulated parts of the subject vehicle (more precisely, the software and hardware of the automated driving system) and the proportion of the road environment that is real, emulated, or simulated, as shown in Fig.~\ref{fig_Antkiewicz}.
Antkiewicz~\emph{et~al.}~\cite{Antkiewicz_2020} pointed out that ``in general, other mixtures of real or simulated hardware and software parts of the [subject vehicle]\footnote{``SV'' in the original text.} under test are also possible.''
As an example, they mentioned the hardware-in-the-loop test method.

\Figure[t!](topskip=0pt, botskip=0pt, midskip=0pt)[width=0.9\columnwidth]{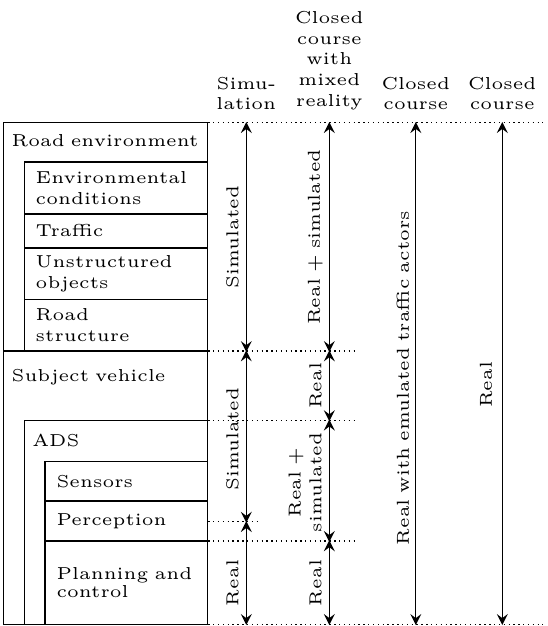}
{Scenario testing modes per test environment (ADS:~automated driving system) based on Antkiewicz~\emph{et~al.}~\cite{Antkiewicz_2020} (adapted to the terms used in this paper: simulated, emulated, and real; additionally, closed-course testing was split into closed-course testing where everything is real and closed-course testing where everything is real but the traffic actors are emulated to illustrate all four testing modes Antkiewicz~\emph{et~al.}~\cite{Antkiewicz_2020} considered).\label{fig_Antkiewicz}}

The authors of the publications cited above classify test benches and test methods in a generic way. 
Thus far, we are not aware of any publication that proposes a method for classifying test bench configurations or a method that considers the elements provided at a specific test bench.
In our opinion, such a classification method is an essential prerequisite for (automatically) assigning test cases to test bench configurations.
Consequently, the requirements identified in Section~\ref{sec_requirements} cannot be fulfilled with the methods described above.
For this reason, we propose a more specific method for classifying test benches and specific test bench configurations in Section~\ref{sec_classification_method}.

\subsection{Test Case Assignment Methods} \label{ssec_related_work_assignment}

In this section, we describe and evaluate the three test case assignment methods known to us.

Böde~\emph{et~al.}~\cite{Bode_2018} dealt with the question of what an optimal (with respect to the resulting test case execution costs) tradeoff between test case execution in simulation and driving tests might look like.
In their method, test case execution in simulation is used to verify the safety properties of the (highly) automated driving function; driving tests are used to validate the simulation models used.
To calculate this tradeoff, Böde~\emph{et~al.}~\cite{Bode_2018} defined a constrained optimization problem for a given overall confidence level and a safety level.
In this calculation, they assumed fixed costs for each simulation run and for each driving test to validate the simulation models.
As a result, the optimal tradeoff between tests in simulation and driving tests is calculated based on the cost of executing a single test in a simulation or a driving test while still ensuring a probability of meeting all requirements.
Unfortunately, Böde~\emph{et~al.}~\cite{Bode_2018} did not list these requirements.
Therefore, they proposed a strategy to split the execution of test cases into simulation and driving tests.
This (binary) splitting is a kind of test case assignment.
However, they emphasized that they did not analyze specific test benches for this splitting but rather examined the general conditions under which such an approach is applicable.
Due to this generic view, the method proposed by Böde~\emph{et~al.}~\cite{Bode_2018} is not suitable for systematically assigning test cases to test bench configurations.

Based on Schuldt~\emph{et~al.}~\cite{Schuldt_2015}, Schuldt~\cite{Schuldt_2017} proposed a method for assigning test cases to X-in-the-loop test methods using his virtual modular test kit.
Unfortunately, these publications are only available in German language.
Therefore, we describe the proposed assignment method in more detail.
Fig.~\ref{fig_assignment_method_Schuldt} shows a schematic illustration of the proposed assignment method. 
The assignment method consists of two steps.

\Figure[t!](topskip=0pt, botskip=0pt, midskip=0pt)[]{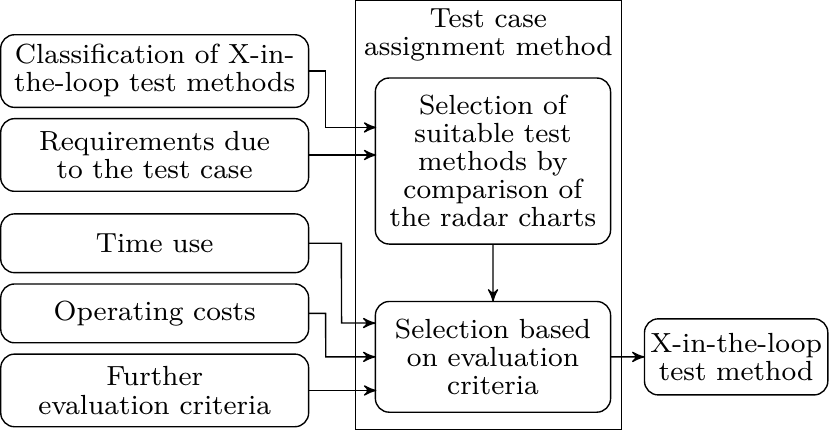}
{Schematic illustration of the method for assigning test cases to X-in-the-loop test methods based on Schuldt~\cite{Schuldt_2017}.\label{fig_assignment_method_Schuldt}}

In the \textbf{first step}, suitable X-in-the-loop test methods for executing the test case are selected based on the classification of X-in-the-loop test methods and the requirements due to the test case.
The classification of the X-in-the-loop test methods is performed using the radar charts described in Section~\ref{ssec_related_work_classification}.
The requirements of the test case define the simulated, emulated, and real elements for executing the test case.
These requirements are also classified using a radar chart.
Schuldt~\cite{Schuldt_2017} did not describe how to determine these requirements. 
Based on the radar charts, X\mbox{-}in-the-loop test methods suitable for executing the test case can be selected by comparing the radar charts.
An X-in-the-loop test method is suitable if the radar chart of the X-in-the-loop test method covers all dimensions of the radar chart of the test case.
Schuldt~\cite{Schuldt_2017} did not further investigate the execution of test cases on a proving ground or on public roads.

In the \textbf{second step}, the optimal X-in-the-loop test method is selected from the remaining X-in-the-loop test methods using an evaluation function (e.g.,~the quality of the provided simulation models, time use, and operating costs).
For this purpose, the radar chart is supplemented by an axis orthogonal to the radar chart on which cost values, consisting of weighted evaluation criteria, are inscribed for every discretization stage. 
Fig.~\ref{fig_weighted_evaluation_function} shows a blank radar chart with the cost values for the dimension of the vehicle dynamics inscribed on the cost-value axis.
It is noteworthy that it is unclear to us why Schuldt~\cite{Schuldt_2017} drew a continuous (blue) line between the different stages. 
Since the stages are discrete, in our opinion, there should be only one cost value at each stage and dimension. 
The line implies that the cost values increase strictly monotonically from the stage simulated to the stage real, which is not the case.
Schuldt~\cite{Schuldt_2017} provided an example of an evaluation function, which is based on weighted cost values.
Schuldt~\cite{Schuldt_2017} did not describe how to determine these cost values.
Moreover, there are inconsistencies in the equations provided by Schuldt~\cite{Schuldt_2017}. 
However, since we do not use these equations in this publication, we explain these inconsistencies in Appendix~\ref{sec_apendixSchuldt}.

\Figure[t!](topskip=0pt, botskip=0pt, midskip=0pt)[]{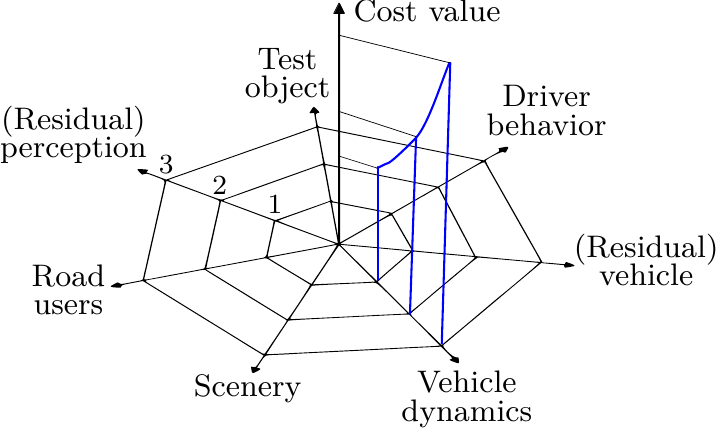}
{Blank radar chart with stages 1~=~simulated, 2~=~emulated, and 3~=~real and the cost values for the dimension of the vehicle dynamics inscribed on the cost-value axis based on Schuldt~\cite{Schuldt_2017}.\label{fig_weighted_evaluation_function}}

In this paragraph, we explain why the method presented by Schuldt~\cite{Schuldt_2017} is not sufficient for assigning test cases to test bench configurations.
According to Schuldt~\cite{Schuldt_2017}, the evaluation of the validity of a simulation model is a challenge, as there were no metrics to evaluate the validity of simulation models at the time of his publication.
Without such an evaluation, it is only possible to evaluate X-in-the-loop test methods based on time use or operating costs \cite{Schuldt_2017}.
However, according to Schuldt~\cite{Schuldt_2017}, the proposed method offers the potential to systematically assign test cases to X-in-the-loop test methods based on model validity when metrics for evaluating the validity of simulation models and emulations become available. 
These metrics are still not available for all possible (types of) simulation models and emulated elements (e.g., Schaermann~\emph{et~al.}~\cite{Schaermann_2017}, Law~\cite{Law_2019}, and Holder~\emph{et~al.}~\cite{Holder_2018}).
Where validity metrics are under development, they are not yet commonly accepted.
However, in our opinion, the validity (domain) of simulation models must not be part of a cost function as proposed by Schuldt~\cite{Schuldt_2017} but is a mandatory prerequisite for generating sufficiently valid test case results.
Only then can the generated test case results be used in the context of evaluating the safe operation of an automated vehicle. 
Furthermore, it is not expedient to assign test cases to test methods since a test method alone does not provide enough information about the performance of a specific test bench configuration that is used to execute the test case. 
A specific test method can be realized on different test benches by different simulation models, emulated elements, and/or real elements with different qualities and credibility (i.e.,~different accuracy, precision, and uncertainty).
An assignment method must be able to consider specific test benches and the test bench configurations they provide.
The aspects mentioned above must be considered when assigning test cases to test bench configurations.
Therefore, the test case assignment method proposed by Schuldt~\cite{Schuldt_2017} is not sufficient for systematically assigning test cases to test bench configurations.

Antkiewicz~\emph{et~al.}~\cite{Antkiewicz_2020} summarized the strengths and weaknesses of the four modes of scenario testing they investigated: closed-course testing with real actors, closed-course testing with robotic surrogate actors, closed-course testing with mixed reality, and simulation testing.
These strengths and weaknesses were determined based on their experience gained from executing six test scenarios with these four testing modes.
Based on this experience, they compared the modes ``using eight criteria: realism, effort and cost, agility, scalability, controllability, accuracy and precision, test safety, and the need to access to [automated driving system]\footnote{``ADS'' in the original text.} internals.''
Finally, Antkiewicz~\emph{et~al.}~\cite{Antkiewicz_2020} provided twelve recommendations for selecting one of these four testing modes.
These recommendations can be used by someone who needs to assign test cases to decide which testing mode should be used to execute a test case. 
Antkiewicz~\emph{et~al.}~\cite{Antkiewicz_2020} emphasized that their ``work uses a slightly consolidated list of test elements, but the main contribution is the comparison of the four testing modes.''
Furthermore, they emphasized that ``an effective verification and validation approach will use a mix of the modes.''
Apart from their recommendations, an (automatic) assignment of test cases to test bench configurations is not considered.
Due to this generic view, the recommendations proposed by Antkiewicz~\emph{et~al.}~\cite{Antkiewicz_2020} are not sufficient for automatically assigning many test cases to test bench configurations.

Thus far, we are not aware of any publication that proposes a method for systematically assigning test cases to test bench configurations, which is mandatory for the effective and efficient execution of a large number of test cases. 
Consequently, the requirements identified in Section~\ref{sec_requirements} cannot be fulfilled with the methods described above.
For this reason, we propose a method for systematically assigning test cases to test bench configurations in Section~\ref{sec_assignment_method}, which allows for an effective and efficient generation of sufficiently valid test case results.

\section{Proposed Method for Classifying Test Benches and Test Bench Configurations} \label{sec_classification_method}

In this section, we propose and describe a method for classifying test benches and test bench configurations.
This method is based on generic functionalities that must be provided at a test bench to execute test cases.
To derive these functionalities systematically, we investigate how an automated vehicle interacts with a driver or a user and its environment during the execution of a test case.
Before conducting this investigation, we describe the relevant terms (driver, user, test case, and automated vehicle) and their associated explanations in the following three paragraphs.

\textbf{Driver and user:} The driver or user is a person who operates the vehicle's human-machine interface. 
The distinction between driver and user is considered meaningful since only a user and not a driver interacts with the vehicle in automated mode (SAE levels~$>$~3~\cite{SAE_J3016_2021}).
Depending on the vehicle's automation level, the user may be able to take over control by driving him- or herself.

\textbf{Test case:} Steimle~\emph{et~al.}~\cite{Steimle_2021} described that in scenario-based test approaches for automated vehicles, a test case consists of a scenario and at least one evaluation criterion. 
According to Bagschik~\emph{et~al.}~\cite{Bagschik_2018b}, based on Schuldt~\cite{Schuldt_2017}, a scenario can be structured by the five\mbox{-}layer model.
According to this model, a scenario can be structured by the road level, traffic infrastructure, temporary manipulation of the road level and traffic infrastructure, movable objects, and the environment. 
Bock~\emph{et~al.}~\cite{Bock_2018} extended this five\mbox{-}layer model with a sixth layer for digital information.
In a later work, Weber~\emph{et~al.}~\cite{Weber_2019} renamed this sixth layer to data and communication.
Scholtes~\emph{et~al.}~\cite{Scholtes_2021} presented the historical development of the six-layer model and described the respective layers in detail.
Fig.~\ref{fig_6layers} shows the six-layer model for structuring scenarios, including examples of the respective layers.
Evaluation criteria may impact the platform used for test case execution, as the data required for test case evaluation must be generated during the execution of the scenario. 
Additionally, it must be possible to evaluate these data.
To generate the necessary data and evaluate them, additional hardware or software may be required that is not directly needed for the actual execution of the scenario but may have a (negative) impact on the test case execution.

\Figure[t!](topskip=0pt, botskip=0pt, midskip=0pt)[scale=0.98]{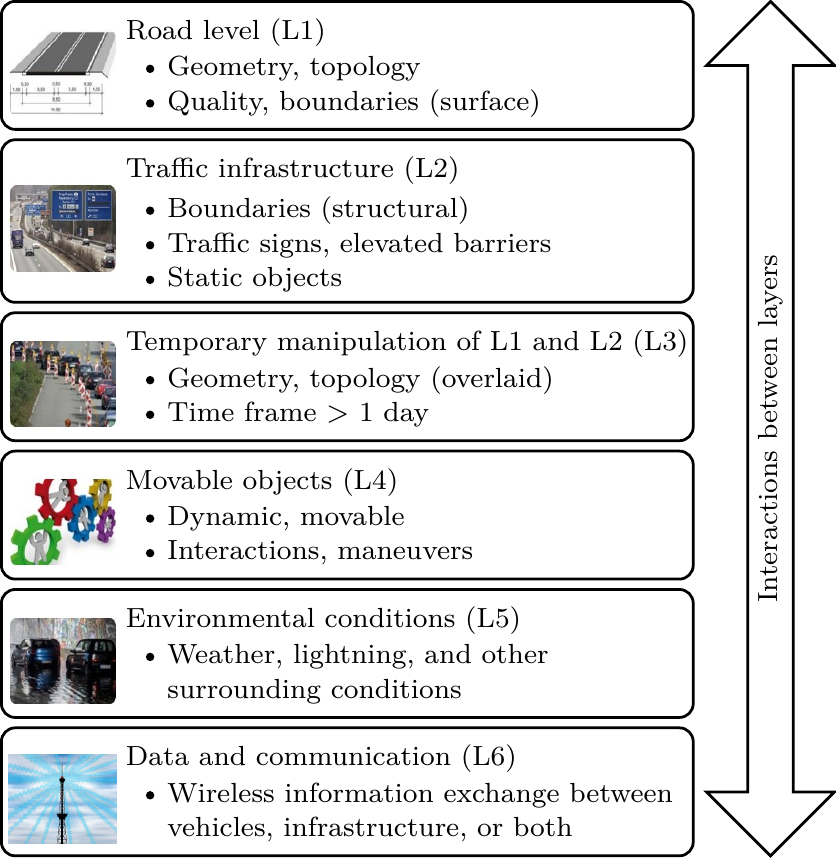}
{Six-layer model for structuring scenarios based on Bagschik~\emph{et~al.}~\cite{Bagschik_2018b} and Weber~\emph{et~al.}~\cite{Weber_2019}.\label{fig_6layers}}

\textbf{Automated vehicle:} An automated vehicle can be functionally described by a functional system architecture.
Such a functional system architecture was developed at the Institute of Control Engineering at TU Braunschweig.
This architecture has been repeatedly discussed and refined (Matthaei and Maurer~\cite{Matthaei_2015}, Matthaei~\cite{Matthaei_2015b}, and Ulbrich~\emph{et~al.}~\cite{Ulbrich_2017}). 
For a complete functional description of an automated vehicle, we extended this architecture with two additional functional blocks: ``vehicle dynamics'' and ``residual vehicle functionalities.''
Fig.~\ref{fig_general_architecture} shows the resulting functional system architecture of an automated vehicle.
The automation system and the different vehicle functionalities are color coded and named. 
``Vehicle dynamics'' describes the movement of the (automated) vehicle in the environment due to the actuation of the actuators.
``Residual vehicle functionalities'' describe functionalities of the components that are not directly required for operating the automation system but are necessary for the complete functional description of an automated vehicle. 
However, residual vehicle functionalities are highly dependent on the specific characteristics of the automation system and the vehicle's characteristics and cannot be generally named. 
For this reason, the functional interdependences between the ``residual vehicle functionalities'' and the specific functional blocks of the ``vehicle functionalities relevant to the automation system'' cannot be generally named, although there are functional interdependences.
Therefore, these connections are drawn only to the vehicle functionalities relevant to the automation system and not to each individual functional block contained therein (see Fig.~\ref{fig_general_architecture}).
Ulbrich~\emph{et~al.}~\cite{Ulbrich_2017} distinguished between environment sensors and vehicle sensors, as shown in Fig.~\ref{fig_general_architecture}.
According to Ulbrich~\emph{et~al.}~\cite{Ulbrich_2017}, environment sensors cover external aspects around the  vehicle (exteroceptive), while vehicle sensors obtain information about the vehicle itself and its internal state (proprioceptive).
Since a more detailed overview and explanation of the other functional blocks is not required for deriving the generic functionalities, we refer interested readers to Matthaei and Maurer~\cite{Matthaei_2015} or Ulbrich~\emph{et~al.}~\cite{Ulbrich_2017}.

\Figure[t!](topskip=0pt, botskip=0pt, midskip=0pt)[width=1.3\columnwidth]{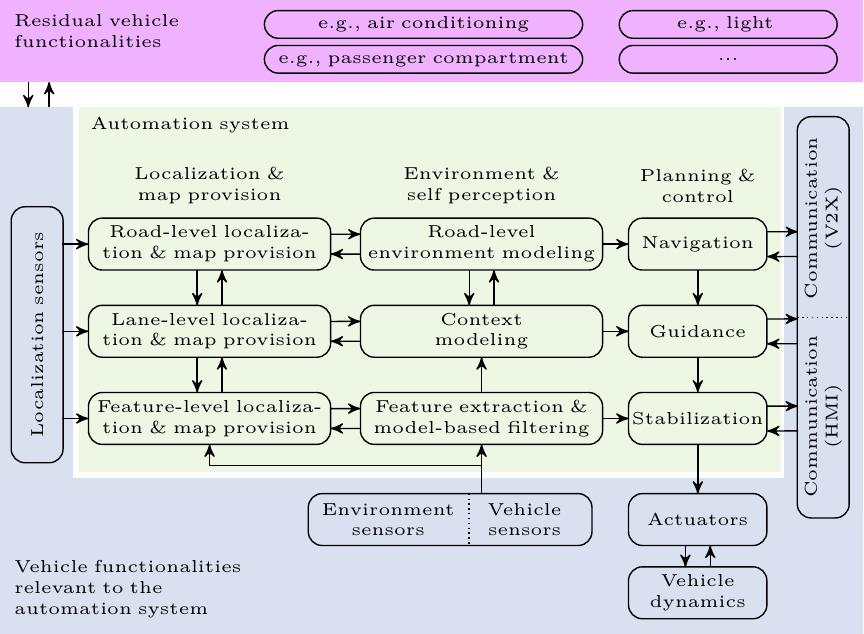}
{Functional system architecture of an automated vehicle (blocks represent functionalities; arrows indicate functional interdependences; HMI:~human-machine interface, V2X: vehicle-to-X) based on Matthaei and Maurer~\cite{Matthaei_2015}.\label{fig_general_architecture}}

As already described, the classification method is based on generic functionalities that must be provided at a test bench to execute test cases.
To derive these functionalities systematically, we investigate the two extreme forms of test benches in the following two paragraphs.
First, we look at a test bench where all elements are real (i.e., a test vehicle).
Second, we look at a test bench where all elements are simulated (i.e., a software-in-the-loop test bench\footnote{
At this point, the term \textit{software-in-the-loop test bench} includes a \textit{model-in-the-loop test bench} where all elements are also simulated. 
In research and industry, it is common to distinguish between these two types of test benches. 
However, since only the two extreme forms of test benches are relevant for deriving the functionalities (i.e., all elements are real or all elements are simulated), this distinction is irrelevant for deriving these functionalities.
}).

\textbf{Test vehicle:} A test vehicle moves in a real environment and can be controlled by a real driver or user; that is, the interfaces of the functional system architecture of an automated vehicle shown in Fig.~\ref{fig_general_architecture} ``communicate'' with the real environment and the real driver or user during the execution of a test case.
Fig.~\ref{fig_architecture_real_sil}~(a) shows the associated architecture.

\Figure[t!](topskip=0pt, botskip=0pt, midskip=0pt)[width=1.75\columnwidth]{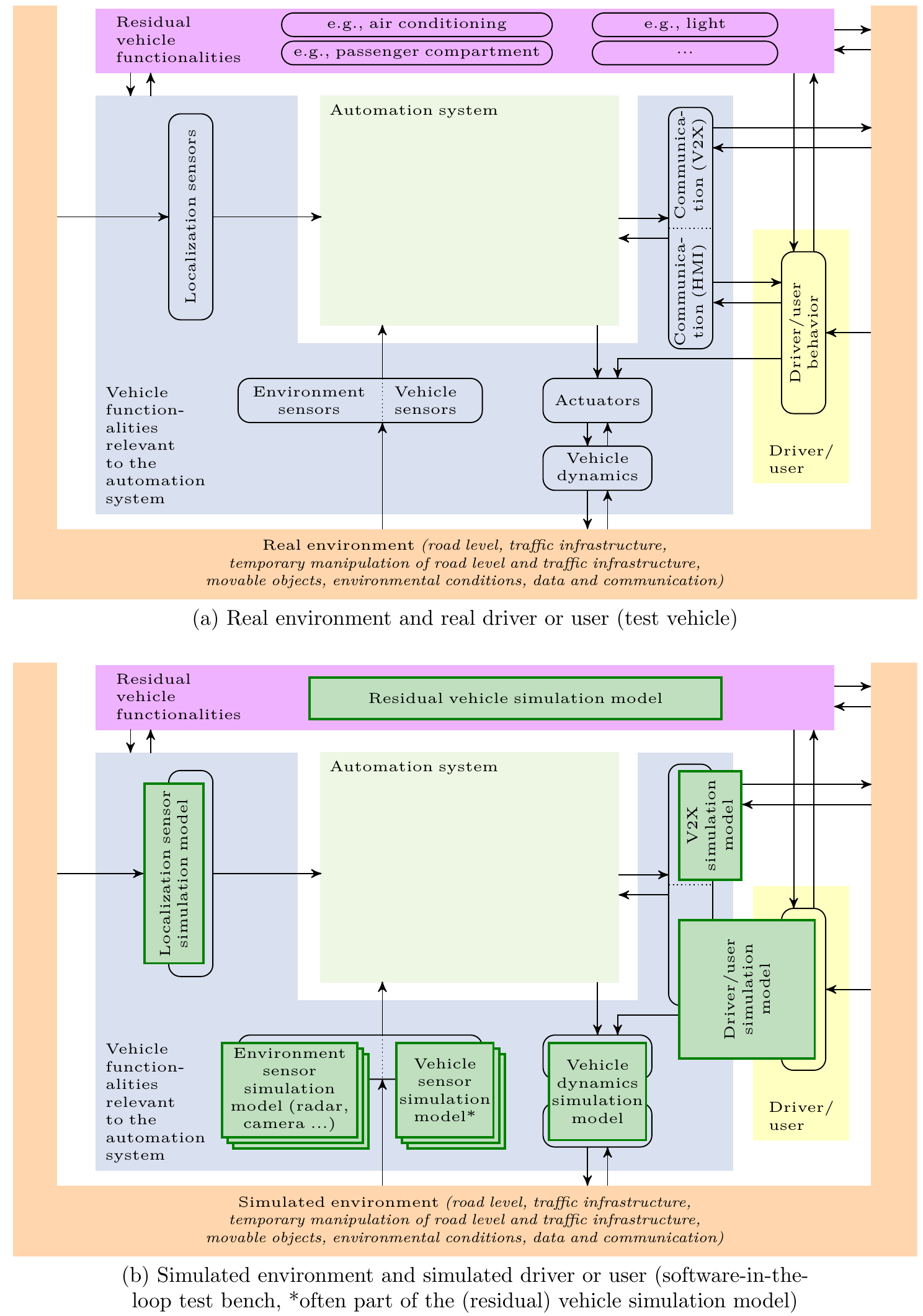}
{Functional system architecture of an automated vehicle embedded in an environment with a driver or user (blocks represent functionalities; arrows indicate functional dependences; HMI: human-machine interface, V2X: vehicle-to-X). The detailed architecture of the automation system is shown in Fig.~\ref{fig_general_architecture}.\label{fig_architecture_real_sil}}

\textbf{Software-in-the-loop test bench:} To operate the automation system at a software-in-the-loop test bench, the interfaces ``communicate'' with a simulated environment and a simulated driver or user. 
Consequently, a simulation environment must provide the same data at the interfaces of the automation system. 
Fig.~\ref{fig_architecture_real_sil}~(b) shows the associated architecture.
Additional functionalities such as simulation control, which are also required to execute a test case with a software-in-the-loop test bench, are not listed because they are not relevant for deriving generic functionalities.

By comparing these two extreme forms of test benches, we derive the generic functionalities, which we call dimensions, that a test bench must provide to execute a test case. 
At a specific test bench, each functionality is implemented by one or more elements.
The derived generic functionalities and a description of each functionality are listed below.\\
	{\color{alizarin}\Pifont{pzd}{\char227}} 
 	\textbf{Scenery}: The dimension of the scenery describes all geospatially stationary elements of the environment~\cite{Ulbrich_2015b}. The scenery is represented by Layer~1 (road level), Layer~2 (traffic infrastructure), and Layer~3 (temporary manipulation of road level and traffic infrastructure) of the six\mbox{-}layer model shown in Fig.~\ref{fig_6layers}. \\
	{\color{alizarin}\Pifont{pzd}{\char227}} 
	\textbf{Movable objects}: The dimension of the movable objects describes objects that move by kinetic energy or are supposed to move within the scenario. The respective objects can be temporarily static.~\cite{Steimle_2021} Movable objects are represented by Layer~4 (movable objects) of the six\mbox{-}layer model shown in Fig.~\ref{fig_6layers}. \\
	{\color{alizarin}\Pifont{pzd}{\char227}} 
	\textbf{Environmental conditions}: The dimension of the environmental conditions describes weather, lighting, and other surrounding conditions. Environmental conditions are represented by Layer~5 (environmental conditions) of the six\mbox{-}layer model shown in Fig.~\ref{fig_6layers}. \\
	{\color{alizarin}\Pifont{pzd}{\char227}} 
	\textbf{V2X communication}: The dimension of V2X communication describes wireless information exchange between vehicles, between infrastructure elements, or between a vehicle and an infrastructure element. V2X communication is represented by Layer~6 (data and communication) of the six\mbox{-}layer model shown in Fig.~\ref{fig_6layers}. \\
	{\color{alizarin}\Pifont{pzd}{\char227}} 
	\textbf{Test object}: The dimension of the test object describes the object to be tested, for example, a component or a system. The test object may contain elements that implement (some of) the functionalities listed below if they are to be tested. For example, the test object may include an environment perception sensor such as a radar sensor. The test object can refer to hardware, software, or a combination of both.\\
	{\color{alizarin}\Pifont{pzd}{\char227}} 
	\textbf{Environment perception sensors}: The dimension of the environment perception sensors describes the characteristics of the sensors used for environment perception.\\
	{\color{alizarin}\Pifont{pzd}{\char227}} 
	\textbf{Localization sensors}: The dimension of the localization sensors describes the characteristics of the sensors used for localization.\\	
	{\color{alizarin}\Pifont{pzd}{\char227}} 
	\textbf{Vehicle dynamics}: The dimension of the vehicle dynamics describes the movement of the (automated) vehicle in the environment due to the actuation of the actuators.\\
	{\color{alizarin}\Pifont{pzd}{\char227}} 
	\textbf{Driver/user behavior}: The dimension of the driver/user behavior describes the actions of the driver or user~--~in response to information received from the environment and the (automated) vehicle~--~for his or her operation of the vehicle's human-machine interface.\\	
	{\color{alizarin}\Pifont{pzd}{\char227}} 
    \textbf{Residual vehicle}: The dimension of the residual vehicle describes the functionalities that are, in addition to the functionalities mentioned above, required to operate the test object during the execution of a test case at a specific test bench (e.g., a rest-bus simulation model, its runtime environment, and the necessary input and output interfaces).  
    However, these additionally required functionalities depend, for example, on the characteristics of the test object, the characteristics of the test case, and the characteristics of the test bench. 
    Therefore, they cannot be named in general.

As mentioned above, these functionalities are implemented at a specific test bench by one or more elements, which can be real, emulated, or simulated.
The proposed definitions of these three stages are listed below.\\
	{\color{alizarin}\Pifont{pzd}{\char227}} 
	\textbf{Real}: 
	If an element is real, it is used as planned in the target vehicle.\\
	{\color{alizarin}\Pifont{pzd}{\char227}} 
	\textbf{Emulated}: 
    If an element is emulated, an element comparable to the target hardware is used that shows functionally equivalent behavior (at its interfaces and with respect to the intended purpose of this element) as the real counterpart it represents.
    Since an emulated element can never exactly encompass all the characteristics of its real counterpart, it is always an abstraction of the real counterpart. 
    Therefore, the emulated element cannot show the entire behavior of the real counterpart but only the reality of interest, which is defined in the development process of the emulated element.\\
	{\color{alizarin}\Pifont{pzd}{\char227}} 
	\textbf{Simulated}: 
	If an element is simulated, a simulation model is used that shows functionally equivalent behavior (at its interfaces and with respect to the intended purpose of this element) as the real counterpart it represents.
    Since a simulation model can never exactly encompass all the characteristics of its real counterpart, it is always an abstraction of the real counterpart.
    Therefore, the simulation model cannot show the entire behavior of the real counterpart but only the reality of interest, which is defined in the development process of the simulation model.

The combination of real, emulated, and simulated elements that implement the generic functionalities listed above can be used to classify test benches and test bench configurations.
Radar charts are suitable for a clear visualization of the classification results due to their multidimensional descriptive character.
Each dimension (i.e., spoke) of the radar chart represents a specific functionality of a test bench.
Each dimension can be divided into three discretization levels (nominal scale) representing real, emulated, and simulated elements.
Fig.~\ref{fig_general_spider} shows the proposed dimensions and the three stages in a blank radar chart.
Examples of real, emulated, and simulated elements in each dimension are listed in Table~\ref{tab_dimensions}.

\begin{table*}[!t]
	\caption{Examples of real, emulated, and simulated elements in each proposed dimension.}
	\label{tab_dimensions}
	\setlength{\tabcolsep}{6pt}
	\begin{center}
		\begin{tabular}{p{70pt} p{83pt} p{157pt} p{140pt}}
			\hline
			\textbf{Dimension}							& \textbf{Real element} 		& \textbf{Emulated element} 					& \textbf{Simulated element} \\
			\hline

			Scenery 									& Real curb or real tree 		& Artificial curb or artificial tree			& Simulation model of a curb or tree  \\

			\rowcolor{platinum}	Movable objects			& Series vehicle  				& Balloon vehicle or crash target  				& Vehicle simulation model  \\

			Environmental \par conditions		& Real rain or real fog				& irrigated lane or artificially generated rain		& Rain simulation model \\

			\rowcolor{platinum} V2X \par communication	& V2X data generated \par by another vehicle & V2X data generated by replicated \par transmitter like development hardware	& V2X data generated by a \par V2X simulation model  \\

			Test object 			& Series electronic \par control unit 			& Development electronic \par control unit				& Software executed in development \par execution environment  \\

			\rowcolor{platinum}	Environment \par perception sensors	& Series radar sensor   		& Development radar sensor 						& Radar simulation model \\

			Localization sensors	 	& Series localization \par sensor   		& Development localization sensor 		& Localization simulation model \\

			\rowcolor{platinum}	Vehicle dynamics		& Vehicle dynamics of \par the series vehicle	& Vehicle dynamics of another vehicle that \par behaves functionally equivalent to the series \par vehicle	& Vehicle dynamics simulation model that \par calculates the movement of the vehicle  \\

			Residual vehicle			& Series vehicle is	used to operate	the test object	& Similar vehicle or similar hardware is used \par to operate the test object & Rest-bus simulation model is used \par to operate the test object \\

			\rowcolor{platinum}	Driver/user behavior	& Test driver 					& Driving robot (that moves the steering wheel) 	& Driver simulation model	\\			
	
			\hline
		\end{tabular}
	\end{center}
\end{table*}

In the following paragraphs, we discuss several general points concerning the proposed classification method.

The functionalities and the three stages mentioned above reflect the current state of our research.
The question remains whether the discretization of the dimensions into three stages is sufficient for an unambiguous classification of many test benches and/or many elements.
A further subdivision of the stages may be necessary to allow a more precise classification of the elements, such as distinguishing between different developmental stages of a real element or different credibility assessment levels of a simulation model as defined in the NASA~7009A standard~\cite{NASA_7009A_2016}.
However, this more detailed discretization does not conflict with the proposed classification method but extends it.

The granularity of the dimensions of the proposed method, visualized using a radar chart, should not be considered rigid; the details can be refined according to one’s needs. 
For example, the dimension of the environment perception sensors can also be refined by the specific environment perception sensors provided at a specific test bench (e.g., radar, lidar, and camera) to allow for a more precise classification.

A test bench can provide various elements in one dimension and stage. 
For example, in the dimension of the vehicle dynamics in the stage simulated, two elements might exist: one \emph{single-track simulation model} and one \emph{double-track simulation model}.

\Figure[t!](topskip=0pt, botskip=0pt, midskip=0pt)[]{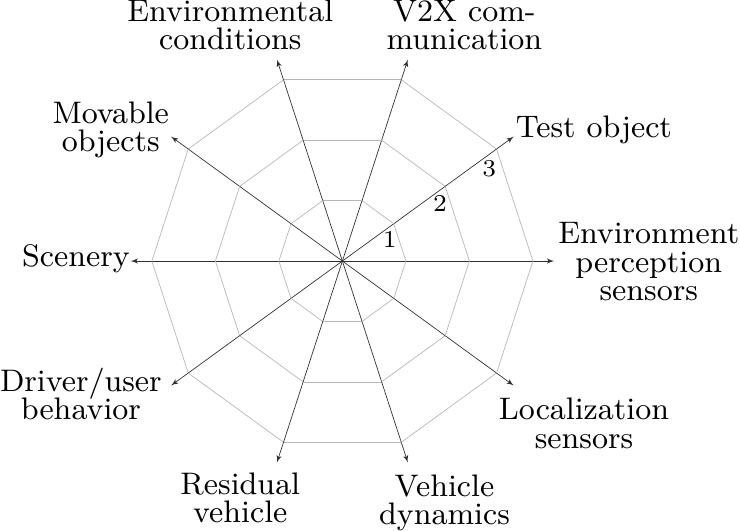}
{Blank radar chart with the proposed dimensions and stages (1~=~simulated, 2~=~emulated, and 3~=~real).\label{fig_general_spider}}

Each element provided at a test bench has specific characteristics, such as its interfaces, validity domain, cost per operating time, and complexity in terms of the necessary execution time.
Ideally, (most) characteristics are defined as (machine-readable) metadata during the development process of the element, for example, a simulation model, and listed in its specification.
These characteristics are not part of this publication; therefore, they are not considered in detail.

As mentioned above, the data required for the test case evaluation must be generated during the execution of the scenario. 
For this reason, an element may require additional functionality in addition to the actual functionality required to execute the scenario (e.g., additional hardware and software). 
This additional functionality is assigned to the respective element when classifying test benches and test bench configurations.

In all dimensions, a combination of simulated, emulated, and real elements within the respective dimension might exist. 
For example, in a test case execution with a vehicle-in-the-loop test bench in the dimension of the movable objects, real vehicles and emulated vehicles (e.g.,~balloon vehicles) might exist.
Both vehicle types are perceived by real environment perception sensors.
In addition, simulated vehicles might exist that are perceived by sensor simulation models.
A detailed description of the vehicle-in-the-loop test method can be found in Berg~\cite{Berg_2014}.
Hallerbach~\emph{et~al.}~\cite{Hallerbach_2018} referred to the concept of transferring perceived simulated objects to an automated driving function as a prototype-in-the-loop approach. In this approach, a real vehicle continuously interacts with a traffic simulation while driving on a real-world proving ground.%

\section{Illustration of the Proposed Classification Method} \label{sec_examples_classification}

In this section, we evaluate the proposed method for classifying test benches and test bench configurations by illustrating various examples. 
First, we classify two example test benches in Section~\ref{ssec_exampleClassificationTestBenches}.
Subsequently, based on these classifications, we derive two test bench configurations in Section~\ref{ssec_exampleClassificationTestBenchConfigurations}.
We use these examples again in Section~\ref{sec_example_assignment}, where we evaluate the proposed test case assignment method by assigning a test case to a test bench configuration.
However, we note that the examples used are intended to illustrate the application of the proposed classification method and are, therefore, intentionally simplified.

\subsection{Classification of Test Benches}  \label{ssec_exampleClassificationTestBenches}

Fig.~\ref{fig_KV_HiL_Test_Bench} visualizes the classification result of a specific hardware-in-the-loop test bench with all elements\footnote{In places where we refer to specific elements provided at a test bench, we write the names of the elements in italics.} provided in each dimension and stage using a radar chart.
This test bench provides two vehicle dynamics models: one \emph{single-track simulation model} and one \emph{double-track simulation model}.
Therefore, there are two elements in the dimension of the vehicle dynamics in the stage simulated. 
To highlight these two simulation models in this publication, they are not drawn exactly on the arrow in Fig.~\ref{fig_KV_HiL_Test_Bench} but are slightly offset.
Furthermore, there is one radar simulation model and one camera simulation model.
Therefore, we refine the dimension of the environment perception sensors with the dimensions of radar and camera to allow a more precise classification.
The dimension of the radar includes in the stage simulated the \emph{radar simulation model}, and the dimension of the camera includes in the stage simulated the \emph{camera simulation model}.
The test object is the \emph{target electronic control unit running the target software}.
For simplicity, this hardware-in-the-loop test bench provides one simulated element in the remaining dimensions.

\Figure[t!](topskip=0pt, botskip=0pt, midskip=0pt)[width=0.95\columnwidth]{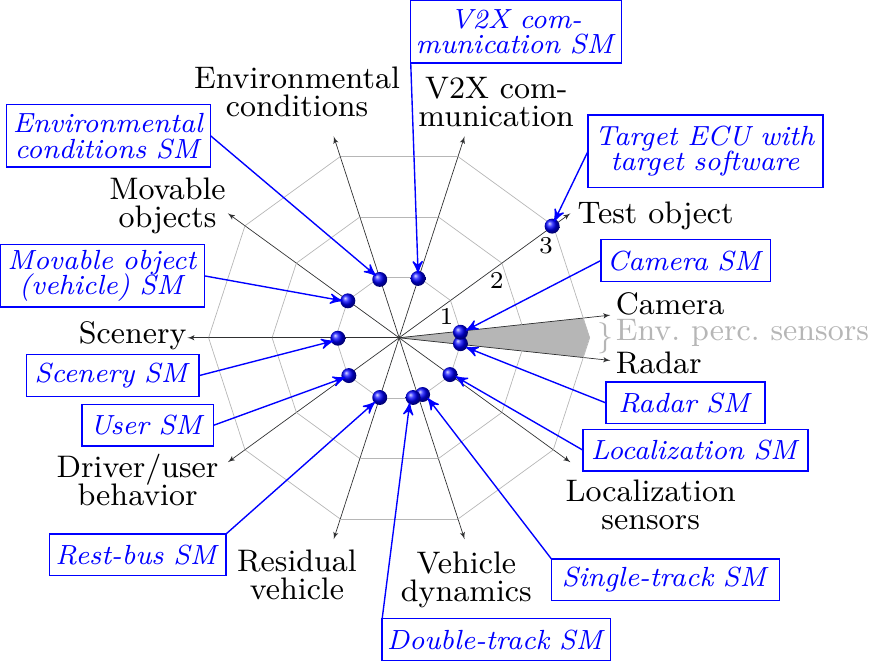}
{Classification result of a specific hardware-in-the-loop test bench visualized using a radar chart with stages 1~=~simulated, 2~=~emulated, and 3~=~real (the blue dots represent the elements provided at this test bench; each element has specific characteristics that are not listed in this figure; Env.~perc.:~environment perception; SM:~simulation model; ECU:~electronic control unit).\label{fig_KV_HiL_Test_Bench}}

Fig.~\ref{fig_KV_TestVehicle} visualizes the classification result of a specific test vehicle with all elements provided in each dimension and stage using a radar chart.
In this example, the test object is the \emph{target electronic control unit running the target software}. 
A \emph{former vehicle} is used to operate the test object (e.g., because the target vehicle is not yet available). 
Therefore, the vehicle dynamics and the residual vehicle are emulated. 
\emph{Pre-production sensors} are used instead of the target sensors (e.g., because the target sensors are not yet available). 
Therefore, the localization sensors, the radar sensor, the camera sensor, and the V2X communication are emulated. 
The vehicle operates on a \emph{proving ground} where the environmental conditions and the scenery are real.
In addition, there is a \emph{real user} inside the vehicle, and as a movable object, there is another \emph{real vehicle}.

As mentioned above, every element illustrated in Figs.~\ref{fig_KV_HiL_Test_Bench} and~\ref{fig_KV_TestVehicle} has specific characteristics, such as its interfaces, validity domain, cost per operating time, and complexity in terms of the necessary execution time.
Therefore, each test bench has specific characteristics. 
Ideally, these characteristics are defined as metadata during the development process of the element, for example, a simulation model, and listed in its specification.
These characteristics are not part of this publication and are still open research questions addressed, for example, in the SET~Level research project~\cite{SL45_Homepage}.

\Figure[t!](topskip=0pt, botskip=0pt, midskip=0pt)[width=0.98\columnwidth]{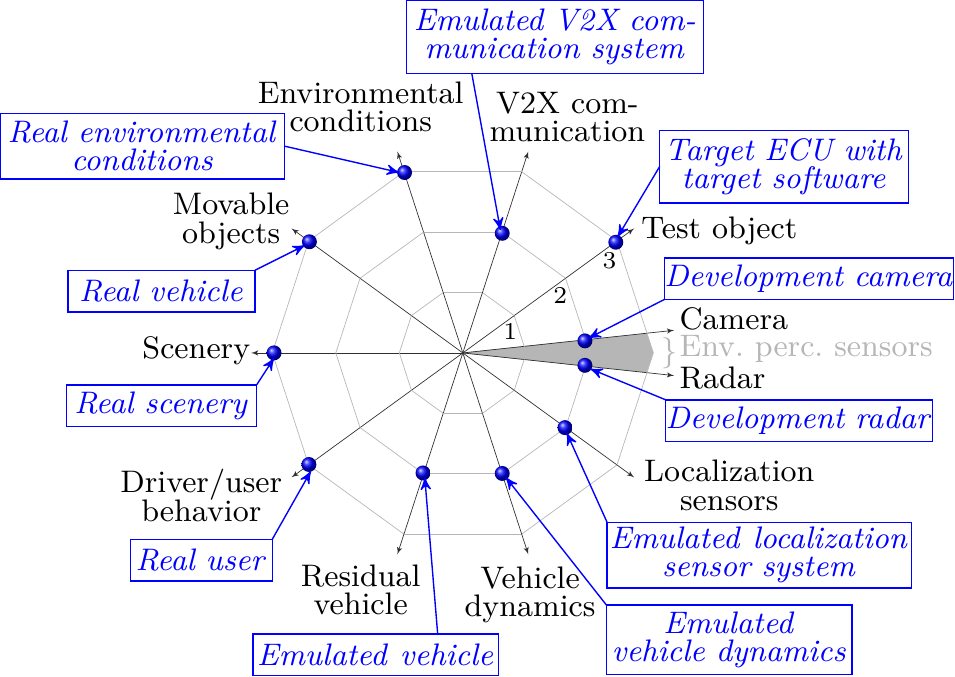}
{Classification result of a specific test vehicle visualized using a radar chart with stages 1~=~simulated, 2~=~emulated, and 3~=~real (the blue dots represent the elements provided at this test bench; each element has specific characteristics that are not listed in this figure; Env.~perc.:~environment perception; ECU:~electronic control unit).\label{fig_KV_TestVehicle}}

\subsection{Classification of Test Bench Configurations} 
\label{ssec_exampleClassificationTestBenchConfigurations}

To execute a test case with a test bench, at least one test bench configuration must be derived from this test bench.
For this purpose, the provided elements (in each dimension of the classified test bench) must be selected and composed.
To derive all test bench configurations provided at a test bench, the elements provided in each dimension must be connected in every meaningful composition.

Based on the classification result of the hardware-in-the-loop test bench visualized in Fig.~\ref{fig_KV_HiL_Test_Bench}, two test bench configurations can be derived.
Fig.~\ref{fig_KV_Example_Test_Bench_Configuration_1} visualizes the first test bench configuration, including the \emph{single-track simulation model} mentioned above, using a radar chart.
This test bench configuration is referred to as \textit{HiL-TBC-1} (HiL:~hardware-in-the-loop; TBC: test bench configuration) in this publication.
The dashed orange line indicates the composition of the elements belonging to this test bench configuration.
The second test bench configuration provided at this test bench, which includes the \emph{double-track simulation model}, can be visualized analogously.
This test bench configuration is referred to as \textit{HiL-TBC-2} in this publication.
As a result, the considered hardware-in-the-loop test bench provides two different test bench configurations, each consisting of a different composition of elements.
In this example, every test bench configuration has a different simulation model in the dimension of the vehicle dynamics.
Therefore, depending on the test bench configuration used to execute a specific test case, its execution may or may not provide sufficiently valid test case results.

\Figure[t!](topskip=0pt, botskip=0pt, midskip=0pt)[width=0.94\columnwidth]{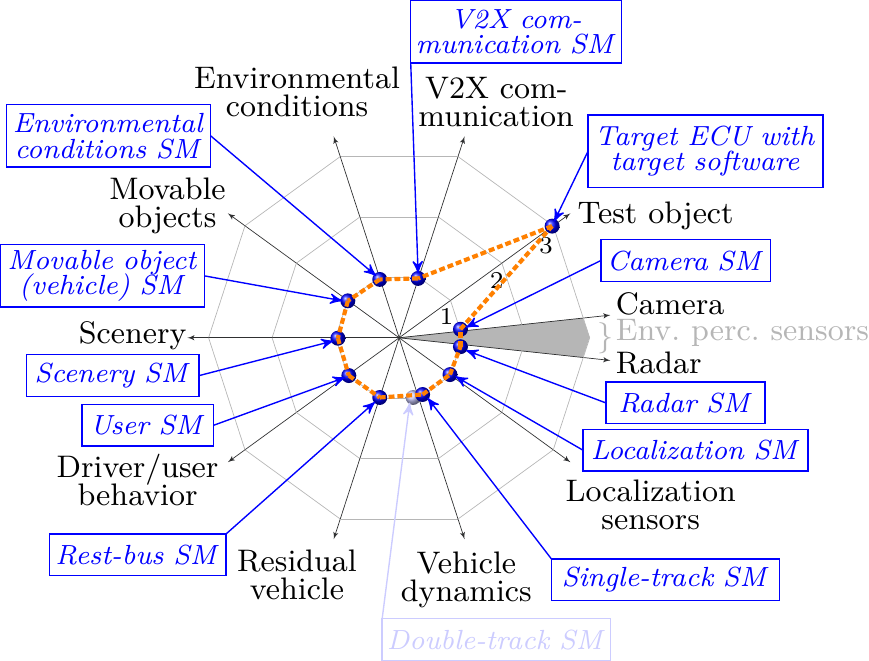}
{Visualization of the test bench configuration \textit{HiL-TBC-1} using a radar chart with stages 1~=~simulated, 2~=~emulated, and 3~=~real (the blue dots represent the elements provided at the corresponding test bench; each element has specific characteristics that are not listed in this figure; the dashed orange line indicates the composition of elements belonging to this test bench configuration; Env.~perc.:~environment perception; SM:~simulation model; ECU:~electronic control unit).\label{fig_KV_Example_Test_Bench_Configuration_1}}

The classification of test bench configurations serves as a basis for the method for systematically assigning test cases to test bench configurations, which is a topic presented in the next section.

\section{Proposed Method for Systematically Assigning Test Cases to Test Bench Configurations} \label{sec_assignment_method} 
In this section, we propose and describe a method for systematically assigning test cases to test bench configurations. 
This method is based on the method for classifying test benches and test bench configurations proposed in Section~\ref{sec_classification_method} and is illustrated schematically as a flow chart in Fig.~\ref{fig_Zuordnungsmethode_neu}. 
The method consists of a structured process that supports, for example, someone who needs to assign test cases to different test bench configurations or execute test cases with sufficient validity, such as a test engineer.
Additionally, this process forms the basis for an implementation, which will allow for the effective, efficient, and automated execution of (extensive) test case catalogs.
We assume that the elements, for example, simulation models, already exist for the assignment of test cases to test bench configurations.
The development of new elements and the further development of existing elements are currently not foreseen when applying the assignment method.
If it is determined during the assignment that there is no test bench that is suitable for operating the test object or no sufficiently valid test bench configuration that can be used to execute a test case, the assignment method must be aborted.
The proposed assignment method consists of a multistep process, which is described in the following paragraphs. 
The respective numbers and designations of the process steps, the input artifacts, and the output artifacts are shown in Fig.~\ref{fig_Zuordnungsmethode_neu}.
To illustrate the proposed test case assignment method with an example, we systematically assign a test case to a test bench configuration according to the proposed method in the next section.

\Figure[t!](topskip=0pt, botskip=0pt, midskip=0pt)[width=0.99\textwidth]{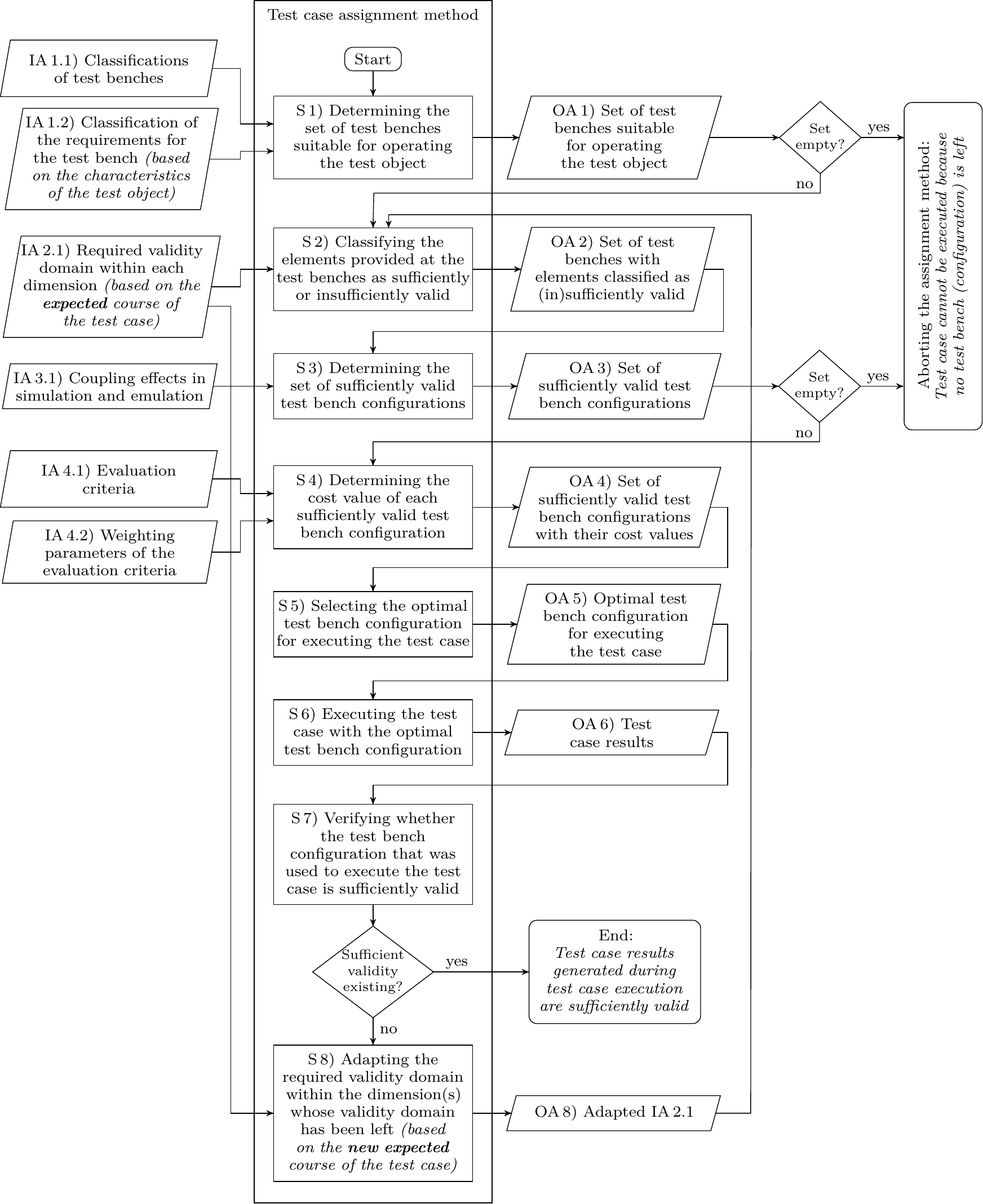}
{Schematic illustration of the proposed method for systematically assigning test cases to test bench configurations as a flow chart (ellipse: start/end; arrow:~process sequence; parallelogram: input/output artifact (IA/OA); rectangle: process step~(S); diamond:~decision).\label{fig_Zuordnungsmethode_neu}}

In \textbf{Step One} (S\,1 in Fig.~\ref{fig_Zuordnungsmethode_neu}), the set of test benches suitable for operating the test object is determined.
The input artifacts (IA) of this process step are the classifications of test benches~(IA\,1.1 in Fig.~\ref{fig_Zuordnungsmethode_neu}) and the classification of the requirements for the test bench, which are based on the characteristics of the test object~(IA\,1.2 in Fig.~\ref{fig_Zuordnungsmethode_neu}).
To create Input Artifact~1.1, all test benches included in the assignment method are classified based on their characteristics according to the classification method described in Section~\ref{sec_classification_method}.
Either these characteristics are part of the specifications of the test benches~--~which includes the specifications of the elements provided at the test benches~--~or they are derived based on an analysis of these test benches.
Each element provided at the considered test bench is assigned to the corresponding dimension and stage.
Each element has specific characteristics, such as its interfaces, validity domain (see also note~(1) described in the following paragraph), cost per operating time, and complexity in terms of the necessary execution time.
The classification results can be visualized using radar charts.
To create Input Artifact~1.2, the requirements for the test bench are classified based on the characteristics of the test object.
Either these characteristics are specified in a dedicated document, such as the test specification, or are derived based on an analysis of the test object.
This classification result can be visualized in the same way using a radar chart.
Then, the set of test benches suitable for operating the test object is determined by comparing the classification results.
A test bench is suitable for operating the test object if all requirements that the test object sets on the test bench are provided at the test bench.
The output artifact (OA) of this process step is the set of (classified) test benches that are suitable for operating the test object~(OA\,1 in Fig.~\ref{fig_Zuordnungsmethode_neu}).
If this set is empty, the assignment method must be aborted (in its current form) since there is no test bench that is suitable for operating the test object.

\textit{Note~(1) on the validity domain of elements}:
Regarding the validity domain of elements, we want to note that determining and evaluating the validity domain of simulation models and emulated elements requires validity metrics.
Currently, metrics that can be used for this determination and evaluation exist only for certain application domains, for example, partly for vehicle dynamics simulation models, and not for all possible (types of) simulation models and emulated elements, for example, sensor simulation models.
Validity metrics that are currently under development are usually not yet commonly accepted (see Section~\ref{sec_related_work}).
The development of validity metrics that can be used to (automatically and objectively) determine and evaluate the validity domain of simulation models and emulated elements is not part of this publication and is still an open research question.
Their development was addressed, for example, in the PEGASUS research project~\cite{PEGASUSMethod_2019} and is currently addressed, for example, in the SET~Level research project~\cite{SL45_Homepage}.
As long as there is no objective knowledge about the validity domains of all elements intended to be used to execute test cases, this process step must be supported by (subjective) expert knowledge to create the necessary input artifact.
However, the expressiveness is limited due to subjective support.
Further research on this input artifact is required to implement and automate this process step in a sufficiently objective manner.
However, the current need for expert knowledge does not conflict with the proposed assignment method.
Rather, it indicates the need for further research to allow for the effective, efficient, and automated execution of (extensive) test case catalogs.

In \textbf{Step Two} (S\,2 in Fig.~\ref{fig_Zuordnungsmethode_neu}), the elements provided at the test benches are classified as sufficiently or insufficiently valid.
The input artifacts of this process step are the set of (classified) test benches that are suitable for operating the test object~(OA\,1 in Fig.~\ref{fig_Zuordnungsmethode_neu}), selected in the previous process step, and the required validity domain within each dimension, which are based on the expected course of the test case~(IA\,2.1 in Fig.~\ref{fig_Zuordnungsmethode_neu}).
Either these required validity domains are already specified in a dedicated document, such as the test case specification, or are derived based on an analysis of the expected course of the test case (see also note~(2) described in the following paragraph).
The elements must provide these validity domains to be sufficiently valid.
Elements that do not provide the required validity domain cannot be used to execute the test case in a sufficiently valid manner.
Subsequently, the elements provided at the remaining test benches are classified as sufficiently or insufficiently valid.
This classification is done by comparing their validity domains with the required validity domain within each dimension (see also note~(3) described in the next but one paragraph).
The output artifact of this process step is the set of test benches with elements classified as sufficiently or insufficiently valid~(OA\,2 in Fig.~\ref{fig_Zuordnungsmethode_neu}).

\textit{Note~(2) on the required validity domains}:
Regarding the required validity domains, we want to note that it may be possible for experts to (manually) specify the required validity domains for a small number of test cases already in the test case specification.
In our opinion, this manual expert-based approach is no longer efficient for extensive test case catalogs that are generated automatically.
Additionally, the expressiveness is limited due to this subjective expert-based support.
Therefore, in the future, it will be necessary to automatically and objectively determine the required validity domains to execute a given test case based on analyzing its expected course.
The development of an appropriate method is not part of this publication and is still an open research question.
Its development was addressed, for example, in the PEGASUS research project~\cite{PEGASUSMethod_2019} and is currently addressed, for example, in the SET~Level research project~\cite{SL45_Homepage}.
As long as there is no possibility to determine these required validity domains automatically and objectively, this process step must be supported by (subjective) expert knowledge to create the necessary input artifact.
However, the expressiveness due to subjective support is limited.
Further research on this input artifact is required to implement and automate this process step in a sufficiently objective manner.
However, the current need for expert knowledge does not conflict with the proposed assignment method.
Rather, it indicates the need for further research to allow the effective, efficient, and automated execution of (extensive) test case catalogs.

\textit{Note~(3) on sufficiently valid elements}:
Regarding sufficiently valid elements, we want to note that, as already described in Step~One, there are currently no metrics to (automatically and objectively) determine and evaluate the validity domain for most (types of) simulation models and emulated elements.
Strictly speaking, for an automated assignment, all simulation models and emulated elements that do not have an objectively determined validity domain would currently have to be classified as not sufficiently valid since there is no objective evidence about their validity domain.

In \textbf{Step Three} (S\,3 in Fig.~\ref{fig_Zuordnungsmethode_neu}), the set of sufficiently valid test bench configurations is determined.
The input artifacts of this process step are the set of test benches with elements classified as sufficiently or insufficiently valid~(OA\,2 in Fig.~\ref{fig_Zuordnungsmethode_neu}), determined in the previous process step, and coupling effects in simulation and emulation~(IA\,3.1 in Fig.~\ref{fig_Zuordnungsmethode_neu}) that have to be considered when coupling elements.
To create all sufficiently valid test bench configurations, the sufficiently valid real, emulated, and simulated elements provided at the respective test bench are coupled in every meaningful composition so that the test case can be executed.
Based on the classification results of the considered test benches, the created test bench configurations can be visualized using radar charts.
When creating test bench configurations, it is essential to consider which elements can be coupled since, for example, no real movable object can be present in simulated scenery.
This consideration includes the interfaces of the elements.
If the interfaces do not match, the elements cannot be coupled. 
There are efforts to standardize interfaces to couple and exchange various simulation models, for example, the Open Simulation Interface~\cite{ASAMOSI_2021}\cite{OSIDocumentation_2021} and the Functional Mock-up Interface~\cite{FMIHomepage_2021}.
Additionally, there are efforts to define complete systems consisting of one or more Functional Mock-up Units, including its parameterization that can be transferred between simulation tools, for example, the System Structure and Parameterization~\cite{SSPHomepage_2021}.
Additionally, all effects resulting from coupling multiple elements are considered when creating the test bench configurations.
In this publication, these effects are referred to as \textit{coupling effects in simulation and emulation}.
This consideration is necessary since this coupling might influence the validity of the test bench configuration to which these coupled elements belong, for example, due to numerical errors or delays in the data exchange between different simulation models~\cite{Kramer_2021} \cite{Frerichs_2021} (see also note~(4) described in the following paragraph).
The output artifact of this process step is the set of (classified) sufficiently valid test bench configurations~(OA\,3 in Fig.~\ref{fig_Zuordnungsmethode_neu}), each of which has a specific performance corresponding to the elements belonging to the respective test bench configuration.
If this set is empty, then the assignment method must be aborted (in its current form) since there is no test bench configuration that can be used to execute the test case.

\textit{Note~(4) on coupling multiple elements}:
Regarding coupling multiple elements, we want to note that there are two ways to consider the influence of this coupling on the validity of the test bench configuration to which these coupled elements belong. 
Ideally, all effects resulting from coupling multiple elements can be determined and validated in advance so that their influence can be directly considered when creating the test bench configuration. 
However, to the best of our knowledge, there are no metrics that can be used to determine and validate these coupling effects (in advance). 
Thus, it may be necessary to (re)validate the test bench configuration to which these coupled elements belong. 
However, there are currently no metrics that can be used to determine and evaluate the validity domain of coupled elements and the test bench configuration to which these coupled elements belong~--~similar to determining and evaluating the validity domain of simulation models and emulated elements.  
The development of validity metrics that can be used to (automatically and objectively) determine and evaluate (in advance) the influence of coupling multiple elements on the validity of the test bench configuration to which these coupled elements belong is not part of this publication and is still an open research question.
Another related open research question addresses the development of validity metrics that can be used to (automatically and objectively) determine and evaluate the validity domain of coupled elements and the test bench configuration to which these coupled elements belong. 
The development of the validity metrics mentioned above is not part of this publication.
The development of validity metrics for simulation models is currently addressed, for example, in the SET Level research project~\cite{SL45_Homepage}. 
To date, we are not aware of any efforts to develop validity metrics for emulated elements.
As long as there is no objective knowledge about the influence of coupling multiple elements on the validity of the test bench configuration or about the validity domain of coupled elements and the test bench configuration to which these coupled elements belong, this process step must be supported by (subjective) expert knowledge. 
However, the expressiveness is limited due to subjective support. 
Therefore, further research on the influence of coupling multiple elements on the validity of the test bench configuration is required to implement and automate this process step in a sufficiently objective manner.
However, the current need for expert knowledge does not conflict with the proposed assignment method.
Rather, it indicates the need for further research to allow the effective, efficient, and automated execution of (extensive) test case catalogs.

In \textbf{Step Four} (S\,4 in Fig.~\ref{fig_Zuordnungsmethode_neu}), the cost value of each sufficiently valid test bench configuration is determined.
The input artifacts of this process step are the set of (classified) sufficiently valid test bench configurations~(OA\,3 in Fig.~\ref{fig_Zuordnungsmethode_neu}), determined in the previous process step, the evaluation criteria~(IA\,4.1 in Fig.~\ref{fig_Zuordnungsmethode_neu}), and the weighting parameters of these evaluation criteria~(IA\,4.2 in Fig.~\ref{fig_Zuordnungsmethode_neu}). 
As long as the evaluation criteria and the values of their weighting parameters do not change, the cost value of the corresponding test bench configuration must be determined only once.
Since the cost value of each sufficiently valid test bench configuration depends on many different influencing factors, it still has to be investigated whether these cost values should be determined quantitatively by a metric or qualitatively by expert knowledge.
A metric for quantitatively determining the cost value of a test bench configuration~(TBC) is described below.
First, the cost value of each sufficiently valid element~(e) provided at the considered test bench~(TB) is determined.
For this purpose, the cost value~($K_{\text{TB,e,c}}$) of every evaluation criterion~(c) (e.g.,~time use or test case execution costs) belonging to a specific element is determined.
These cost values are then weighted according to the weighting parameters~($a_{\text{c}}$) of the corresponding evaluation criterion and summed.
As a result, the cost value~($K_{\text{TB,e}}$) of each sufficiently valid element provided at a specific test bench is determined.
Eqs.~(\ref{eq_costValue_Element}) and~(\ref{eq_costValue_addOn}) show the corresponding formulas.
Subsequently, the cost value~($K_{\text{TBC}}$) of each sufficiently valid test bench configuration is determined. 
For this purpose, the cost values~($K_{\text{TB,e}}$) of all elements belonging to the respective test bench configuration are summed.
Eq.~(\ref{eq_costValueTestBenchConfiguration}) shows the corresponding formula.
The output artifact of this process step is the set of sufficiently valid test bench configurations with their cost values~(OA\,4 in Fig.~\ref{fig_Zuordnungsmethode_neu}).

\begin{equation}
\label{eq_costValue_Element}
K_{\text{TB,e}} = \sum_{\text{all evaluation criteria}} a_{\text{c}} \cdot K_{\text{TB,e,c}}
\end{equation}

\begin{equation}
	\label{eq_costValue_addOn}
	\sum_{\text{all evaluation criteria}} a_{\text{c}} = 1
	\vspace{0.4cm}
\end{equation}

\begin{equation}
\label{eq_costValueTestBenchConfiguration}
K_{\text{TBC}} = \sum_{\text{all elements belonging to the TBC}} K_{\text{TB,e}}
\end{equation}

In \textbf{Step Five}~(S\,5 in Fig.~\ref{fig_Zuordnungsmethode_neu}), the optimal test bench configuration for executing the test case is selected.
The input artifact of this process step is the set of sufficiently valid test bench configurations with their cost values~(OA\,4 in Fig.~\ref{fig_Zuordnungsmethode_neu}), determined in the previous process step.
The optimal test bench configuration for executing the test case ($\text{TBC}_{\text{optimal}}$) is selected based on the lowest cost value of all sufficiently valid test bench configurations according to Eq.~(\ref{eq_TBCToBeUsed}).
The output artifact of this process step is the optimal test bench configuration for executing the test case~(OA\,5 in Fig.~\ref{fig_Zuordnungsmethode_neu}).

\begin{equation}
\label{eq_TBCToBeUsed}
\text{TBC}_{\text{optimal}} = \argmin_{\text{all TBCs}} (K_{\text{TBC}})
\end{equation}

In \textbf{Step Six}~(S\,6 in Fig.~\ref{fig_Zuordnungsmethode_neu}), the test case is executed.
The input artifact of this process step is the optimal test bench configuration for executing the test case~(OA\,5 in Fig.~\ref{fig_Zuordnungsmethode_neu}), selected in the previous process step.
The test case is executed with this test bench configuration.
Thereby, the test case results are generated.
The output artifact of this process step consists of these test case results~(OA\,6 in Fig.~\ref{fig_Zuordnungsmethode_neu}).

The actual course of the test case usually cannot be predicted with certainty before the test case is executed due to the open behavior of the automated vehicle. 
Therefore, it is also impossible to predict with certainty whether the test bench configuration that was used to execute the test case and the test case results are sufficiently valid.
If the automated vehicle behaves differently than expected, elements may leave their validity domain.
As a result, the test bench configuration that was used to execute the test case and the generated test case results may no longer be sufficiently valid\footnote{This problem could be reduced by already considering all possible (even extreme) contingencies in Step~Two (at least as far as it is possible). 
However, this contradicts the requirement that for the most efficient test case execution, for example, simulation models should not be more detailed than necessary.}.
Consider the following example, where a particular test case is executed with a software-in-the-loop test bench.
At the beginning of the test case, the automated vehicle drives on the right lane of a two-lane highway behind another vehicle.
At a specific point in time, the other vehicle starts braking. 
It is assumed that the automated vehicle has two options: it can either brake and follow the other vehicle or it can perform a lane change.
Imagine that when the vehicle dynamics simulation model was selected, it was assumed that the automated vehicle will also brake.
This assumption places low demands on the lateral acceleration validity domain of the vehicle dynamics simulation model.
If the automated vehicle then performs a very dynamic lane change during the test case execution, the vehicle dynamics simulation model might have left its validity domain.

Therefore, in \textbf{Step Seven}~(S\,7 in Fig.~\ref{fig_Zuordnungsmethode_neu}), it is verified whether the test bench configuration that was used to execute the test case~--~selected in Step Five~--~is sufficiently valid.
For this purpose, it is verified~--~either simultaneously with executing the test case or after executing the test case based on recorded data~--~whether elements belonging to this test bench configuration have left their validity domain during the execution of the test case.
If no element has left its validity domain, the test case has been executed sufficiently valid.
Consequently, the test case results are also sufficiently valid, and they can be used for their intended purpose, for example, to assess the ego vehicle's safe behavior.
In this case, the test case assignment method is terminated.
In the case of failed test cases, the test object must be adapted accordingly (the error(s) must be corrected), and relevant test cases must be executed again according to the test case assignment method.
If an element has left its validity domain, this element is no longer sufficiently valid.
Consequently, the test bench configuration that includes this element is no longer sufficiently valid.
As a result, the generated test case results are also not sufficiently valid and cannot be used for their intended purpose, for example, to assess the ego vehicle's safe behavior.
In this case, the assignment method must be continued in Step Eight.

In \textbf{Step Eight} (S\,8 in Fig.~\ref{fig_Zuordnungsmethode_neu}), the required validity domain within the dimension(s) whose validity domain has been left is adapted.
The input artifact of this process step is the required validity domain within each dimension, which was determined based on the (previously) expected course of the test case in Step Two~(IA\,2.1 in Fig.~\ref{fig_Zuordnungsmethode_neu}).
In this artifact, the required validity domain within the dimension(s) whose validity domain has been left is adapted.
For this purpose, the new expected course of the test case is analyzed, and the required validity domains are adapted accordingly.
The output artifact of this process step consists of the adapted required validity domain within each dimension~(OA\,8 in Fig.~\ref{fig_Zuordnungsmethode_neu}), which was created by adapting the Input Artifact~2.1.
Based on this adaptation, the process must be repeated from Step Two on.
%

\section{Illustration of the Proposed Test Case Assignment Method} \label{sec_example_assignment}

In this section, we evaluate the proposed test case assignment method by assigning a test case to a test bench configuration as an example. 
Note that the example is intentionally simplified.
In reality, the input artifacts and the implementation of the process steps are not that simple.
In this example, however, we focus on illustrating the assignment method, not on assigning extensive test case catalogs.

As an example application, we consider a highway chauffeur (SAE level~3~\cite{SAE_J3016_2021}).
The test object is the \emph{target electronic control unit running the software of the highway chauffeur}.
The test case considered consists of the concrete scenario named ``cut-in vehicle'' and the evaluation criterion that the distance between the ego vehicle and all other objects is greater than \SI{0}{\meter} (no collision with other objects) throughout the scenario.
A graphical overview of the initial scene of the corresponding functional scenario is illustrated in Fig.~\ref{fig_szenario_merging_vehicle}. 
Additionally, the behavior of the vehicles during the scenario is indicated by the dotted arrows.
An excerpt of the parameter values of the considered concrete scenario is listed in Table~\ref{tab_parameterValues}.
Furthermore, it is assumed that the following three test benches are available to execute the test case (the elements provided at each of the test benches are listed later):

\begin{itemize}
	\item Software-in-the-loop test bench 
	\item Hardware-in-the-loop test bench
	\item Test vehicle
\end{itemize}

In the next sections, we apply the proposed method for systematically assigning test cases to test bench configurations to generate sufficiently valid test case results.
The respective numbers and designations of the process steps, the input artifacts, and the output artifacts are shown in Fig.~\ref{fig_Zuordnungsmethode_neu}.

\subsubsection*{Applying Step One: Determining the set of test benches suitable for operating the test object (see S\,1 in Fig.~\ref{fig_Zuordnungsmethode_neu})} \label{ssec_step_1}

In this process step, the set of test benches suitable for operating the test object is selected.
For this purpose, the test benches included in the assignment method must be classified based on their characteristics according to the classification method presented in Section~\ref{sec_classification_method}.
These classification results represent Input Artifact~1.1 (IA\,1.1 in Fig.~\ref{fig_Zuordnungsmethode_neu}).
Additionally, the requirements for the test bench, which are based on the characteristics of the test object, are classified.
This classification result represents Input Artifact~1.2 (IA\,1.2 in Fig.~\ref{fig_Zuordnungsmethode_neu}).

To create Input Artifact~1.1, each element provided at the three test benches available in this example must be assigned to the corresponding dimension and stage.
Fig.~\ref{fig_spider_SiL_test_bench_example} visualizes the classification result of the software-in-the-loop test bench using a radar chart.
Figs.~\ref{fig_KV_HiL_Test_Bench} and~\ref{fig_KV_TestVehicle}, which are already presented in Section~\ref{sec_examples_classification}, visualize the classification results of the hardware-in-the-loop test bench and the test vehicle using radar charts.
Each element shown in these figures has specific characteristics, such as its validity domain.
As mentioned above, currently, there are no (commonly accepted) metrics to determine and evaluate the validity domains of all possible (types of) simulation models and emulated elements.
Therefore, as described in the explanation of this process step in Section~\ref{sec_assignment_method}, this process step must currently be supported by expert knowledge to determine the validity domains of the elements provided at the available test benches.
For simplicity, the validity domains are given only for the two vehicle dynamics simulation models provided at the hardware-in-the-loop test bench (see also Fig.~\ref{fig_KV_HiL_Test_Bench}) -- the \emph{single-track simulation model} (e.g., a linear single-track model with a linear tire model) and the \emph{double-track simulation model} (e.g., a double-track model with a nonlinear tire model) -- since these two simulation models will play a significant role in the further course of the example.
It is assumed that the \emph{single-track simulation model} is sufficiently valid in the lateral acceleration range of \SIrange{-3}{3}{\meter \per \second\squared} and in the longitudinal acceleration range of \SIrange{-6}{6}{\meter \per \second\squared}.
Furthermore, the \emph{double-track simulation model} is sufficiently valid in the lateral acceleration range of \SIrange{-8}{8}{\meter \per \second\squared} and in the longitudinal acceleration range of \SIrange{-8}{8}{\meter \per \second\squared}.
Note that these values are given as examples to demonstrate the assignment method.
The classification results of these three test benches represent Input Artifact~1.1.

\Figure[t!](topskip=0pt, botskip=0pt, midskip=0pt)[]{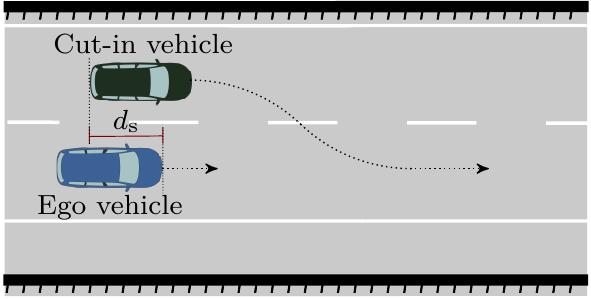}
{Initial scene of the scenario named ``cut-in vehicle''; the dotted arrows indicate the behavior of the vehicles during the scenario (not drawn to scale).\label{fig_szenario_merging_vehicle}}

\begin{table}[t]
	\caption{Excerpt of the parameter values of the considered concrete scenario; $v_{\text{E}}$:~speed of ego vehicle; $v_{\text{C}}$:~speed of cut-in vehicle; $d_\text{s}$:~distance between the front bumper of the ego vehicle and the rear bumper of the cut-in vehicle at the start of the scenario; $d_\text{sm}$:~distance between the front bumper of the ego vehicle and the rear bumper of the cut-in vehicle at which the lane change maneuver starts (see also Fig.~\ref{fig_szenario_merging_vehicle}).}
	\label{tab_parameterValues}
	\setlength{\tabcolsep}{6pt}
	\begin{center}
		\begin{tabular}{l}
			\hline

			$v_{\text{E}} = \SI{120}{\kilo\meter\per\hour}$ \\		
			
			\rowcolor{platinum}	$v_{\text{C}} = \SI{130}{\kilo\meter\per\hour}$ \\
			
			$d_{\text{s}} = \SI{-3}{\meter}$ \\
			
			\rowcolor{platinum}	$d_{\text{sm}} = \SI{20}{\meter}$ \\

			\hline
		\end{tabular}
	\end{center}
\end{table}

\Figure[t!](topskip=0pt, botskip=0pt, midskip=0pt)[width=0.98\columnwidth]{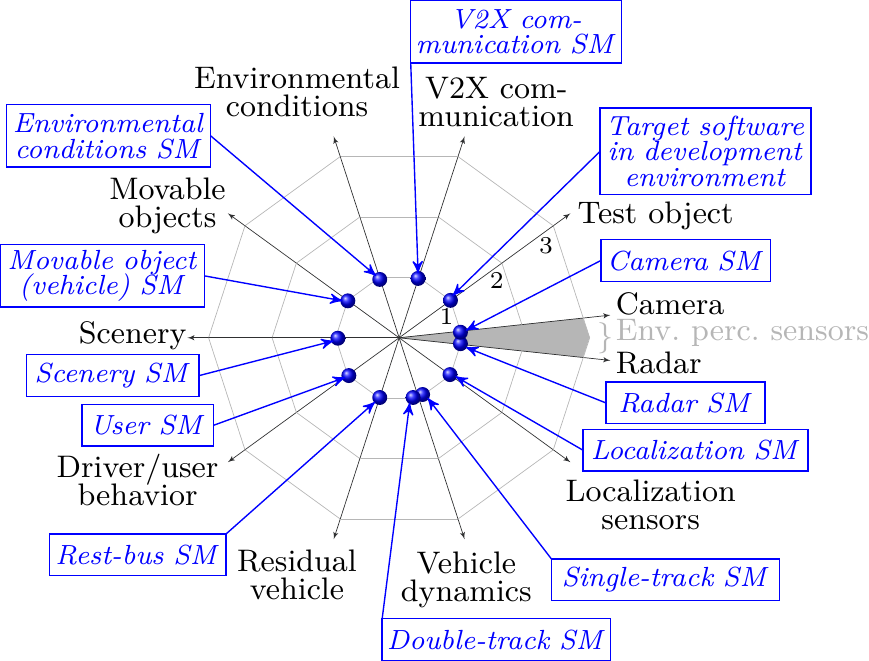}
{Classification result of the software-in-the-loop test bench visualized using a radar chart with stages 1~=~simulated, 2~=~emulated, and 3~=~real (the blue dots represent the elements provided at this test bench; each element has specific characteristics that are not listed in this figure; Env.~perc.:~environment perception; SM:~simulation model).\label{fig_spider_SiL_test_bench_example}}

To create Input Artifact~1.2, the requirements for the test bench, which are based on the characteristics of the test object, must be classified.
In this example, the test object is the \emph{target electronic control unit running the software of the highway chauffeur}. 
It follows that the real electronic control unit must be connectable to the test bench.
From the electronic control unit's point of view, it is irrelevant whether the other dimensions of the test bench are real, emulated, or simulated.
Fig.~\ref{fig_spider_testobject_example} visualizes this classification result using a radar chart, which represents Input Artifact~1.2.
The orange area represents the area of elements usable for operating the test object.

\Figure[t!](topskip=0pt, botskip=0pt, midskip=0pt)[]{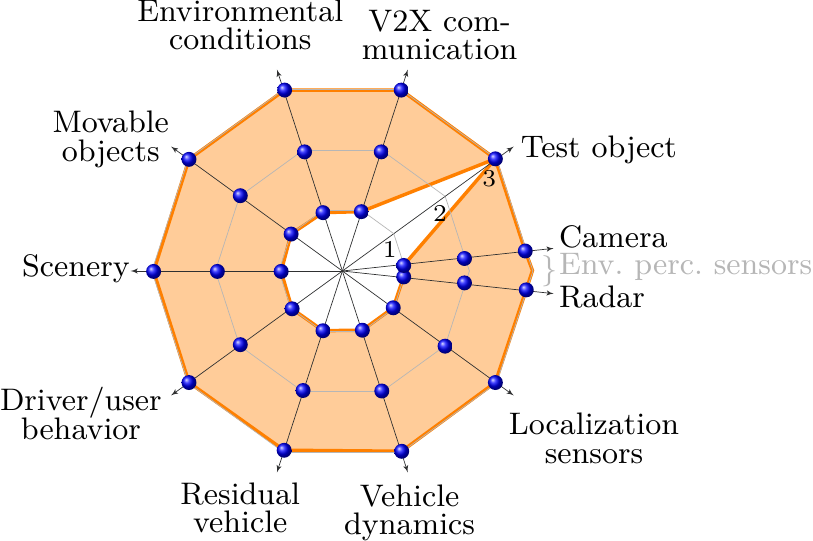}
{Classification result of the requirements for the test bench resulting from the test object visualized using a radar chart with stages 1~=~simulated, 2~=~emulated, and 3~=~real (the orange area represents the area of elements usable for operating the test object; the blue dots indicate these possible elements; Env.~perc.:~environment perception).\label{fig_spider_testobject_example}}

By comparing the classification results of the available test benches visualized in Figs.~\ref{fig_KV_HiL_Test_Bench}, \ref{fig_KV_TestVehicle}, and~\ref{fig_spider_SiL_test_bench_example}~(IA\,1.1 in Fig.~\ref{fig_Zuordnungsmethode_neu}) with the 
classification result of the requirements for the test bench visualized in Fig.~\ref{fig_spider_testobject_example}~(IA\,1.2 in Fig.~\ref{fig_Zuordnungsmethode_neu}), test benches suitable for operating the test object are determined.
For this purpose, all requirements that the test object sets on the test bench must be provided at the test bench.
Therefore, the software-in-the-loop test bench cannot be used to operate the real test object since no real test object can be connected to this test bench.
The hardware-in-the-loop test bench and the test vehicle can be used to operate the real test object and are, therefore, suitable for operating the target electronic control unit.
These two (classified) test benches~--~the hardware-in-the-loop test bench and the test vehicle~--~represent Output Artifact~1 (OA\,1 in Fig.~\ref{fig_Zuordnungsmethode_neu}).
This process step may seem trivial but is indispensable for an automated assignment of test cases.

\subsubsection*{Applying Step Two: Classifying the elements provided at the test benches as sufficiently or insufficiently valid (see S\,2 in Fig.~\ref{fig_Zuordnungsmethode_neu})} \label{ssec_step_2}

In this process step, the elements provided at the remaining test benches~--~the hardware-in-the-loop test bench and the test vehicle, which were determined in the previous process step (S\,1 in Fig.~\ref{fig_Zuordnungsmethode_neu})~--~are classified as sufficiently or insufficiently valid.
For this purpose, the required validity domain within each dimension must be determined.
These required validity domains represent Input Artifact~2.1 (IA\,2.1 in Fig.~\ref{fig_Zuordnungsmethode_neu}).

To create Input Artifact~2.1, the expected course of the test case must be analyzed since, in this example, the required validity domains have not yet been specified in another document.
The expected course of the test case is described below.
The cut-in vehicle passes the ego vehicle, which is driven by the activated highway chauffeur.
The cut-in vehicle starts changing lanes as soon as the distance between the ego vehicle and the cut-in vehicle is \SI{20}{\meter} (see Fig.~\ref{fig_szenario_merging_vehicle}). 
The ego vehicle detects the cut-in vehicle, and it is expected that the ego vehicle brakes slightly to increase the distance accordingly. 
Based on analyzing this expected course, the required validity domain is determined within each dimension.
As mentioned above, there is currently no way to automatically analyze the test case and automatically determine the required validity domains based on this analysis. 
Therefore, this process step must currently be supported by expert knowledge.
For simplicity, the validity domain is given only for the dimension of the vehicle dynamics since this dimension will play a significant role in the further course of the example.
It is assumed that the longitudinal acceleration must be sufficiently valid in the range of \SIrange{-2}{0.5}{\meter \per \second\squared} since the ego vehicle will brake slightly.
Furthermore, it is assumed that the lateral acceleration must be sufficiently valid in the range of \SIrange{-0.5}{0.5}{\meter \per \second\squared} since the ego vehicle will continue to drive and will not steer.
The determined required validity domain within each dimension represents Input Artifact~2.1.

By comparing the validity domains of the elements provided at the remaining test benches~(OA\,1 in Fig.~\ref{fig_Zuordnungsmethode_neu})~--~the hardware-in-the-loop test bench and the test vehicle~--~with the required validity domain within each dimension~(IA\,2.1 in Fig.~\ref{fig_Zuordnungsmethode_neu}), the elements provided at these test benches are classified as sufficiently or insufficiently valid.
Concerning the dimension of the vehicle dynamics, based on the validity domain determined by expert knowledge in Step One, the \emph{single-track simulation model} and the \emph{double-track simulation model}~--~both are provided at the hardware-in-the-loop test bench~--~are classified as sufficiently valid.
Furthermore, it is assumed that all other elements provided at the hardware-in-the-loop test bench and the elements provided at the test vehicle are classified as sufficiently valid.
These two classified test benches with the elements classified as sufficiently or insufficiently valid represent Output Artifact~2 (OA\,2 in Fig.~\ref{fig_Zuordnungsmethode_neu}).

\subsubsection*{Applying Step Three: Determining the set of sufficiently valid test bench configurations (see S\,3 in Fig.~\ref{fig_Zuordnungsmethode_neu})} \label{ssec_step_3}

In this process step, the set of sufficiently valid test bench configurations is determined.
For this purpose, the sufficiently valid elements provided in each dimension and stage of the remaining test benches~--~that were determined in the previous process step (S\,2 in Fig.~\ref{fig_Zuordnungsmethode_neu})~--~must be coupled in every meaningful composition so that the test case can be executed.
For this composition, all effects resulting from coupling these elements must be considered since this coupling might influence the validity of the test bench configuration to which these elements belong.
These coupling effects represent Input Artifact~3.1 (IA\,3.1 in Fig.~\ref{fig_Zuordnungsmethode_neu}).
For simplicity, in this example, it is assumed that there are no coupling effects that affect the validity of the test bench configuration.

The available hardware-in-the-loop test bench provides the \emph{single-track simulation model} and the \emph{double-track simulation model} in the dimension of the vehicle dynamics (see also Fig.~\ref{fig_KV_HiL_Test_Bench}).
As already mentioned, one sufficiently valid element is assumed in all other dimensions.
These compositions result in two sufficiently valid test bench configurations for the hardware-in-the-loop test bench. 
Fig.~\ref{fig_KV_Example_Test_Bench_Configuration_1} in Section~\ref{sec_examples_classification} visualizes the test bench configuration, which includes the \emph{single-track simulation model}, using a radar chart.
This test bench configuration is referred to as \textit{HiL-TBC-1} (HiL:~hardware-in-the-loop; TBC:~test bench configuration) in this publication. 
The second test bench configuration, which includes the \emph{double-track simulation model}, is visualized in the same way.
This second test bench configuration is referred to as \textit{HiL-TBC-2} in this publication.
For the test vehicle, only one composition of the provided elements is possible, which is referred to as \textit{TV\hbox{-}TBC\hbox{-}1} (TV:~test vehicle) in this publication.
These three sufficiently valid test bench configurations represent Output Artifact~3 (OA\,3 in Fig.~\ref{fig_Zuordnungsmethode_neu}).

\subsubsection*{Applying Step Four: Determining the cost value of each sufficiently valid test bench configuration (see S\,4 in Fig.~\ref{fig_Zuordnungsmethode_neu})} \label{ssec_step_4}

In this process step, the cost value of each sufficiently valid test bench configuration~--~which was determined in the previous process step (S\,3 in Fig.~\ref{fig_Zuordnungsmethode_neu})~--~is determined.
For this purpose, the evaluation criteria representing Input Artifact~4.1 (IA\,4.1 in Fig.~\ref{fig_Zuordnungsmethode_neu}) and the weighting parameters of these evaluation criteria representing Input Artifact~4.2 (IA\,4.2 in Fig.~\ref{fig_Zuordnungsmethode_neu}) must be defined.

In this example, the evaluation criteria include the time use (including preparation and follow-up time) and the test case execution costs. 
Both evaluation criteria are equally weighted.
As previously described, it still has to be investigated whether a metric is required to quantify the cost values of the sufficiently valid test bench configurations or whether expert knowledge is better suited for the qualitative determination of these cost values. 
In this example, the cost values are qualitatively determined by expert knowledge.
The cost values given below are determined as examples to illustrate the assignment method.
An exact determination of these cost values and the associated scale is unnecessary to illustrate the assignment method; therefore, it is not explained in detail.

For the test bench configuration \textit{HiL-TBC-1}, which includes the \emph{single-track simulation model}, a cost value of three results from expert knowledge since little preparation and follow-up time is necessary to execute the test case, execution in real time is required, and the test case execution is cost-effective.

For the test bench configuration \textit{HiL-TBC-2}, which includes the \emph{double-track simulation model}, a cost value of four results from expert knowledge since the test bench, other than the dimension of the vehicle dynamics, has the same rating as the configuration that includes the \emph{single-track simulation model}.
However, the \emph{double-track simulation model} is somewhat more computationally intensive and, therefore, demands higher energy costs and more calculation time.

For the test bench configuration \textit{TV-TBC-1}, a cost value of eight results from expert knowledge since a high preparation time and medium follow-up time are necessary to execute the test case, execution in real time is required, and test case execution is relatively expensive.

These three sufficiently valid test bench configurations with their cost values represent Output Artifact~4 (OA\,4 in Fig.~\ref{fig_Zuordnungsmethode_neu}).

\subsubsection*{Applying Step Five: Selecting the optimal test bench configuration for executing the test case (see S\,5 in Fig.~\ref{fig_Zuordnungsmethode_neu})} \label{ssec_step_5}

In this process step, the optimal test bench configuration for executing the test case is selected.
For this purpose, the cost values of the sufficiently valid test bench configurations~--~that were determined in the previous process step (S\,4 in Fig.~\ref{fig_Zuordnungsmethode_neu})~--~must be compared.
The lowest cost value determines the optimal test bench configuration for executing the test case. 
In this example, comparing the cost values of the three sufficiently valid test bench configurations results in using the test bench configuration \textit{HiL-TBC-1} to execute the test case.
This test bench configuration is provided at the hardware-in-the-loop test bench and includes the \emph{single-track simulation model}.
This test bench configuration, which represents Output Artifact~5 (OA\,5 in Fig.~\ref{fig_Zuordnungsmethode_neu}), is visualized in Fig.~\ref{fig_KV_Example_Test_Bench_Configuration_1}.

\subsubsection*{Applying Step Six: Executing the test case with the optimal test bench configuration (see S\,6 in Fig.~\ref{fig_Zuordnungsmethode_neu})} \label{ssec_step_6}

In this process step, the test case, consisting of the scenario named ``cut-in vehicle'' (see Fig.~\ref{fig_szenario_merging_vehicle}) and the evaluation criterion that the distance between the ego vehicle and all other objects is greater than \SI{0}{\meter} throughout the scenario, is executed with the test bench configuration called \textit{HiL-TBC-1}, which was selected in the previous process step (S\,5 in Fig.~\ref{fig_Zuordnungsmethode_neu}).
Thereby, the test case results are generated.
These test case results represent Output Artifact~6 (OA\,6 in Fig.~\ref{fig_Zuordnungsmethode_neu}).

\subsubsection*{Applying Step Seven: Verifying whether the test bench configuration that was used to execute the test case is sufficiently valid (see S\,7 in Fig.~\ref{fig_Zuordnungsmethode_neu})} \label{ssec_step_7}

In this process step, it is verified whether the test bench configuration that was used to execute the test case~--~which was selected in Step Five (S\,5 in Fig.~\ref{fig_Zuordnungsmethode_neu})~--~is sufficiently valid.
For this purpose, it must be verified~--~either simultaneously with executing the test case or after executing the test case based on recorded data~--~whether elements that are part of this test bench configuration have left their validity domain during the execution of the test case.

In this example, the automated vehicle does not follow the lane (as expected in Step Two) but makes a (dynamic) lane change, and the maximum lateral acceleration is \SI{3.5}{\meter \per \second\squared}.
Note that this value is given as an example to demonstrate the assignment method.
By comparing the corresponding validity domain of the \emph{single-track simulation model} (\SIrange{-3}{3}{\meter \per \second\squared}) with this maximum lateral acceleration, it can be stated that the \emph{single-track simulation model} has left its validity domain and is, therefore, no longer sufficiently valid.
Consequently, the test bench configuration that includes this \emph{single-track simulation model} is no longer sufficiently valid.
As a result, the generated test case results are not sufficiently valid.
Insufficiently valid test case results cannot be used for their intended purpose, for example, to assess the ego vehicle's safe behavior.
Therefore, in this example, the assignment method must be continued in Step Eight.

\subsubsection*{Applying Step Eight: Adapting the required validity domain within the dimension(s) whose validity domain has been left (see S\,8 in Fig.~\ref{fig_Zuordnungsmethode_neu})} \label{ssec_step_8}

In this process step, the required validity domain within the dimension(s) whose validity domain has been left is adapted.
The input artifact of this process step is the required validity domain within each dimension, which was determined based on the (previously) expected course of the test case~(IA\,2.1 in Fig.~\ref{fig_Zuordnungsmethode_neu}) in Step Two.
Concerning this artifact, the required validity domain within the dimension(s) whose validity domain has been left must be adapted based on the new expected course of the test case.

To adapt the Input Artifact~2.1 (IA\,2.1 in Fig.~\ref{fig_Zuordnungsmethode_neu}), the new expected course of the test case must be analyzed.
The new expected course of the test case is described below.
The cut-in vehicle passes the ego vehicle, which is driven by the activated highway chauffeur.
The cut-in vehicle starts changing lanes as soon as the distance between the ego vehicle and the cut-in vehicle is \SI{20}{\meter} (see Fig.~\ref{fig_szenario_merging_vehicle}). 
The ego vehicle detects the cut-in vehicle and performs a lane change.
Based on analyzing this new expected course of the test case, the required validity domains are adapted.
In this example, the \emph{single-track simulation model}, which belongs to the dimension of the vehicle dynamics, has left its validity domain.
Therefore, the required validity domain in the dimension of the vehicle dynamics must be adapted.
Instead of the previous validity domain of the lateral acceleration, which is in the range of \SIrange{-0.5}{0.5}{\meter \per \second\squared}, this validity domain must be adapted so that it is now in the range of \SIrange{-3.5}{3.5}{\meter \per \second\squared}.
Furthermore, it is assumed that the required validity domain within all other dimensions remains the same.
This adapted required validity domain within each dimension that is created by adapting the Input Artifact~2.1 represents Output Artifact~8 (OA\,8 in Fig.~\ref{fig_Zuordnungsmethode_neu}).

\subsubsection*{Continue in applying Step Two: Classifying the elements provided at the test benches as sufficiently or insufficiently valid (see S\,2 in Fig.~\ref{fig_Zuordnungsmethode_neu})} \label{ssec_step_2_new}

In this process step, the elements provided at the remaining test benches~--~the hardware-in-the-loop test bench and the test vehicle that were already determined in Step One (S\,1 in Fig.~\ref{fig_Zuordnungsmethode_neu})~--~are classified as sufficiently or insufficiently valid. 
This classification is based on the adapted Input Artifact~2.1 (adapted IA\,2.1 in Fig.~\ref{fig_Zuordnungsmethode_neu}), which was created in Step~Eight (S\,8 in Fig.~\ref{fig_Zuordnungsmethode_neu}).
By comparing the validity domains of the elements provided at the remaining test benches (OA\,1 in Fig.~\ref{fig_Zuordnungsmethode_neu}) with the adapted required validity domain within each dimension (adapted IA\,2.1 in Fig.~\ref{fig_Zuordnungsmethode_neu}), the elements are classified as sufficiently or insufficiently valid.

Concerning the adapted dimension of the vehicle dynamics, the \emph{single-track simulation model}~--~provided at the hardware-in-the-loop test bench~--~is sufficiently valid in the lateral acceleration range of \SIrange{-3}{3}{\meter \per \second\squared}.
Therefore, based on the adapted validity domain within this dimension, this simulation model must be classified as insufficiently valid.
The \emph{double-track simulation model}~--~also provided at the hardware-in-the-loop test bench~--~is sufficiently valid in the lateral acceleration range of \SIrange{-8}{8}{\meter \per \second\squared}.
Consequently, it is still classified as sufficiently valid.
Since the required validity domains within the other dimensions have not changed, all other elements provided at the hardware-in-the-loop test bench continue to be sufficiently valid.
Additionally, it is assumed that all elements provided at the test vehicle continue to be sufficiently valid.
These two classified test benches with the elements they provide that are classified as sufficiently or insufficiently valid represent the adapted Output Artifact~2 (OA\,2 in Fig.~\ref{fig_Zuordnungsmethode_neu}).

Within the next process steps, it is determined that the test bench configuration \textit{HiL-TBC-2}, which is provided at the hardware-in-the-loop test bench and includes the \emph{double-track simulation model}, is the optimal test bench configuration for executing the test case again (see also note~(5) described in the following paragraph).
Then, the test case is executed again with this newly determined test bench configuration.
During this execution, no element has left its validity domain.
Therefore, the test case results are sufficiently valid and can be used for their intended purpose, for example, to assess the ego vehicle's safe behavior.
Hence, the method is terminated.

\textit{Note~(5)}:
If the hardware-in-the-loop test bench would not provide the \emph{double-track simulation model} in the dimension of the vehicle dynamics, this test bench would not provide another sufficiently valid test bench configuration in this example.
Therefore, this test bench cannot be used to execute the test case again.
Then, within the next process steps, it is determined that the test bench configuration \textit{TV-TBC-1}, which is provided at the test vehicle, will be used to execute the test case again. %

\section{Evaluation of the Requirements Set for the Classification and Test Case Assignment Method} \label{sec_req_evaluation}

We created the method for classifying test benches and test bench configurations and the method for systematically assigning test cases to test bench configurations based on the requirements listed in Section~\ref{sec_requirements}.
In this section, we evaluate the fulfillment of these requirements.
As already mentioned, the methods will be implemented in future work.
In doing so, the requirements listed in Section~\ref{sec_requirements} will continue to be considered.

\subsection{Classification Method} \label{ssec_req_eval_classificationMethod}

In this subsection, we evaluate the fulfillment of the requirements on the classification
method.

We derived the criteria used for classifying test benches and test bench configurations systematically based on generic functionalities that a test bench must provide to execute test cases (see evaluation of Req.~C3 below).
We referred to these criteria as dimensions and stages.
Since these criteria are used to systematically classify test benches and test bench configurations, we consider Req.~C1a to be fulfilled.

The elements provided at a specific test bench can be objectively assigned to a dimension and stage. 
However, there is still a need for research concerning the objective specification of the performance of these elements (e.g., concerning their validity domains). 
While developing the classification method, we considered that the performance of the elements provided at a test bench can be objectively specified in the future.
We will continue to consider this requirement as soon as it is possible to specify the performance of the elements objectively.
By coupling elements provided at a test bench to create one or more test bench configurations, it will also be possible to describe the performance of these test bench configurations.
Therefore, we consider Req.~C1b to be partly fulfilled.

We developed the classification method in such a way that the elements provided at a specific test bench can be assigned to the corresponding dimensions and stages. 
Thus, these elements can be considered in the classification of test benches and test bench configurations, even though there is still a need for research to objectively describe the performance of these elements in detail (e.g., concerning their validity domains).
Therefore, we consider Req.~C2 to be fulfilled.

To systematically derive the criteria used for the classification, we investigated how an automated vehicle interacts with a driver or a user and its environment during the execution of a test case.
Based on this investigation, we systematically derived these criteria (i.e., the dimensions and stages).
For this reason, we consider Req.~C3 to be fulfilled.

According to discussions with researchers in the field and the feedback we received, the proposed classification method is as simple and easily understandable as possible.
However, we cannot conclusively evaluate whether this requirement is sufficiently fulfilled for other engineers without extensive feedback from the community.
Therefore, we consider Req.~C4 to be fulfilled to the degree to which we can evaluate it without extensive feedback from the community.

While developing the classification method, we paid particular attention to the fact that the method can be automated.
However, since the method is not yet automated, we cannot make any reliable statement about the fulfillment of this requirement (Req.~C5). 
Therefore, we will consider this requirement in our future work.

We paid particular attention to the fact that the classification results are machine-readable as we developed the classification method. 
However, since the method is not yet automated, we cannot make any reliable statement about the fulfillment of this requirement (Req.~C6). 
Therefore, we will consider this requirement in our future work.

We proposed visualizing the classification results using radar charts. 
According to the researches we discussed with and the feedback we received, this form of visualizing the classification results is intuitive.
However, we cannot conclusively evaluate whether this requirement is sufficiently fulfilled for other engineers without extensive feedback from the community.
Therefore, we consider Req.~C7 to be fulfilled to the degree to which we can evaluate it without extensive feedback from the community.

As described above, we cannot conclusively evaluate the fulfillment of some requirements we have identified without feedback from the community. 
Therefore, we welcome any feedback from the community on the proposed classification method.

\subsection{Test Case Assignment Method} \label{ssec_req_eval_assignmentMethod}

In this subsection, we evaluate the fulfillment of the requirements on the test case assignment
method.

The proposed test case assignment method is based on a process we developed.
We described this process and visualized it schematically for a quick overview.
By applying this process, test cases can be systematically assigned to test bench configurations. 
Therefore, we consider Req.~A1a to be fulfilled.

The assignment itself necessitates the described input artifacts. 
Test cases can be objectively assigned to test bench configurations if the input artifact data are objective.
As mentioned above, there is still a need for research in this regard.
At the moment, we cannot make any reliable statement about the fulfillment of this requirement (Req.~A1b). 
Therefore, we will consider this requirement in our future work.

While developing the test case assignment method, we paid particular attention to the fact that it is possible to generate sufficiently valid test case results.
For this purpose, it is verified after executing the test case whether the generated test case results (and the used test bench configuration) are sufficiently valid. If the test case results are not sufficiently valid, there is a feedback loop for selecting a new test bench configuration so that the test case can be executed again with this configuration.
However, the actual generation of sufficiently valid test case results depends on the metrics used to determine and evaluate the validity domain of the test bench configuration used to execute the test case.
In turn, this validity domain, depends, for example, on the validity domains of the elements it contains.
Currently, metrics that can be used for this determination and evaluation do not exist for all possible (types of) simulation models and emulated elements.
The development of these metrics is not part of this publication.
However, the proposed test case assignment method considers generating sufficiently valid test case results, but the results themselves do not depend on the method.
Therefore, we consider Req.~A2a to be fulfilled.

The most efficient test bench configuration can be selected based on cost values that are determined for each sufficiently valid test bench configuration in applying the test case assignment method.
However, the calculation of the cost values depends on the respective evaluation criteria and their weighting parameters.
Therefore, someone who applies the test case assignment method can determine the criteria that are important to him or her.
Based on these cost values, the most efficient test bench configuration is then selected.
Since the method allows efficient generation of sufficiently valid test case results, we consider Req.~A2b to be fulfilled.

According to discussions with researchers in the field and the feedback we received, the proposed test case assignment method is as simple and easily understandable as possible.
However, we cannot conclusively evaluate whether this requirement is sufficiently fulfilled for other engineers without extensive feedback from the community.
Therefore, we consider Req.~A3 to be fulfilled to the degree to which we can evaluate it without extensive feedback from the community.

While developing the test case assignment method, we paid particular attention to the fact that the method can be automated.
However, since the method is not yet automated, we cannot make any reliable statement about the fulfillment of this requirement (Req.~A4). 
Therefore, we will consider this requirement in our future work.

As already mentioned, there is still a need for research on some input artifacts of the proposed test case assignment method to fully automate this method. 
This research is not part of this publication. 
Therefore, we paid particular attention to the fact that the test case assignment method can be applied with expert knowledge.
Additionally, by applying expert knowledge, we demonstrated the test case assignment method with an example.
Therefore, we consider Req.~A5 to be fulfilled.

As described above, we cannot conclusively evaluate the fulfillment of some requirements we have identified without feedback from the community. 
Therefore, we welcome any feedback from the community on the proposed test case assignment method.
%

\section{Conclusion and Outlook} \label{sec_conclusion}

In this publication, we presented a method for classifying test benches and test bench configurations.
Based on this classification method, we presented a method for systematically assigning test cases to test bench configurations to generate sufficiently valid test case results.
The classification method is based on generic functionalities that a test bench must provide to execute test cases.
To derive these functionalities systematically, we investigated how an automated vehicle interacts with a driver or a user and its environment during the execution of a test case.
Furthermore, we proposed a description of each functionality.
At a specific test bench, each functionality is implemented by one or more elements, which can be real, emulated, or simulated.
We have also proposed a description of these three stages.
To visualize the classification results, we proposed using radar charts, where every dimension represents a specific functionality that can be real, emulated, or simulated.
To evaluate the classification method, we provided various examples of classified test benches and test bench configurations.

Based on the proposed classification method, we presented a method for systematically assigning test cases to test bench configurations, which allows for the effective and efficient generation of sufficiently valid test case results.
The proposed assignment method consists of a multistep process that supports, for example, someone who needs to assign test cases to different test benches or execute test cases with sufficient validity, such as a test engineer.
We described all the process steps, their necessary input artifacts, and their output artifacts.
This process forms the basis for an implementation, which will allow for the effective, efficient, and automated execution of extensive test case catalogs.
To illustrate the proposed test case assignment method, we systematically assigned a test case to a test bench configuration according to the proposed method as an example.

Additionally, we highlighted some of the current limitations in using simulation models and emulated elements for test case execution.
Further research on the following research questions is required to implement and automate the proposed test case assignment method in a sufficiently objective manner.
Explanations regarding these questions are provided in Section~\ref{sec_assignment_method}.

\begin{itemize}
    \item The first identified open research question is how to (automatically and objectively) determine and evaluate the validity domain of simulation models and emulated elements.
    \item The second identified open research question is how to (automatically and objectively) determine and evaluate (in advance) the influence of coupling multiple elements on the validity of the test bench configuration to which these coupled elements belong.
    \item The third identified related open research question is how to (automatically and objectively) determine and evaluate the validity domain of coupled elements and the test bench configuration to which these coupled elements belong.
    \item The fourth identified open research question is how to (automatically and objectively) determine the required validity domains to execute a given test case based on analyzing its expected course.
\end{itemize}

Further research on these questions is required to implement and automate the proposed test case assignment method in a sufficiently objective manner, which will allow the automated execution of extensive test case catalogs.
Meanwhile, the proposed assignment method needs to be supported by expert knowledge.
However, the current need for expert support does not conflict with the proposed assignment method.
Rather, it indicates the need for further research to allow the effective, efficient, and automated execution of extensive test case catalogs.

We continue our activities, for example, in the SET Level research project~\cite{SL45_Homepage}~--~a project of the PEGASUS family~\cite{PEGASUSMethod_2019}~--~that is funded by the German Federal Ministry for Economic Affairs and Energy.

Regarding the proposed methods, we welcome any feedback from the community.%

\appendices

\section{Note on the equations provided by Schuldt}  \label{sec_apendixSchuldt}

As already mentioned in Section~\ref{sec_related_work}, there are inconsistencies in the equations Schuldt~\cite{Schuldt_2017} provided. 
In this section, we explain these inconsistencies.

As an example of an evaluation function, the overall cost value $G$ is the sum of the weighted cost values [see Eqs.~(\ref{eq_evaluation_function_1}) and~(\ref{eq_evaluation_function_2})]. 
$a_{\text{n}}$ is the weighting factor for evaluation criterion~n (e.g.,~the time use or operating cost), and $k_{\text{ijn}}$ is the cost value of dimension~j (e.g.,~the test object or vehicle dynamics) at stage~i (1~=~simulated, 2~=~emulated, 3~=~real) for evaluation criterion~n.
Schuldt~\cite{Schuldt_2017} did not describe how to determine these cost values.
Moreover, Schuldt~\cite{Schuldt_2017} did not explain the meanings of N and J, so we have two comprehension problems with Eqs.~(\ref{eq_evaluation_function_1}) and~(\ref{eq_evaluation_function_2}).
If N represents the number of evaluation criteria and J the number of dimensions, as we assume, n should either start at 1 or go up to $\text{N}-1$, and j should either start at 1 or go up to $\text{J}-1$.
However, there is a second problem with using n and j in the equations.
In the equations, n and j represent numbers.
In the text, however, n and j represent the evaluation criteria and dimensions in textual form.
Therefore, in our opinion, n and j should be replaced with other symbols, and additionally, there must be an assignment of the numbers to the respective evaluation criteria and dimensions.

\begin{equation}
	\label{eq_evaluation_function_1}
	G = \sum_{\text{n}=0}^{\text{N}}a_{\text{n}} \cdot \left( \sum_{\text{j}=0}^{\text{J}}k_{\text{ijn}} \right)
\end{equation}

\begin{equation}
	\label{eq_evaluation_function_2}
	\sum_{\text{n}=0}^{\text{N}}a_{\text{n}} = 1
	\vspace{0.4cm}
\end{equation}

\section*{Acknowledgment}

We want to thank Till Menzel (TU Braunschweig) for the productive discussions and the feedback on our approaches. 
Additionally, we want to thank Ulrich Eberle, Dirk Frerichs, Karl-Michael Hahn, and Christoph Thiem (Opel Automobile, Stellantis) for their peer review of this publication prior to its submission to IEEE \textit{Access}.

\bibliographystyle{IEEEtran}%
\bibliography{literature/literature}%

\begin{thebibliography}{10}
\providecommand{\url}[1]{#1}
\csname url@samestyle\endcsname
\providecommand{\newblock}{\relax}
\providecommand{\bibinfo}[2]{#2}
\providecommand{\BIBentrySTDinterwordspacing}{\spaceskip=0pt\relax}
\providecommand{\BIBentryALTinterwordstretchfactor}{4}
\providecommand{\BIBentryALTinterwordspacing}{\spaceskip=\fontdimen2\font plus
\BIBentryALTinterwordstretchfactor\fontdimen3\font minus
  \fontdimen4\font\relax}
\providecommand{\BIBforeignlanguage}[2]{{%
\expandafter\ifx\csname l@#1\endcsname\relax
\typeout{** WARNING: IEEEtran.bst: No hyphenation pattern has been}%
\typeout{** loaded for the language `#1'. Using the pattern for}%
\typeout{** the default language instead.}%
\else
\language=\csname l@#1\endcsname
\fi
#2}}
\providecommand{\BIBdecl}{\relax}
\BIBdecl

\bibitem{SAE_J3016_2021}
\emph{{Taxonomy and Definitions for Terms Related to Driving Automation Systems
  for On-Road Motor Vehicles}}, {SAE International} Std. J3016, 2021.

\bibitem{Riedmaier_2020}
S.~Riedmaier, T.~Ponn, D.~Ludwig, B.~Schick, and F.~Diermeyer, ``{Survey on
  scenario-based safety assessment of automated vehicles},'' \emph{{IEEE
  Access}}, vol.~8, pp. 87\,456--87\,477, 2020,
  \href{https://doi.org/10.1109/ACCESS.2020.2993730}{doi:
  10.1109/ACCESS.2020.2993730}.

\bibitem{Wachenfeld_2016b}
W.~Wachenfeld and H.~Winner, ``{The release of autonomous vehicles},'' in
  \emph{{Autonomous Driving: Technical, Legal and Social Aspects}}, M.~Maurer,
  J.~C. Gerdes, B.~Lenz, and H.~Winner, Eds.\hskip 1em plus 0.5em minus
  0.4em\relax Springer, 2016, pp. 425--449.

\bibitem{Stellet_2020}
J.~E. Stellet, M.~Woehrle, T.~Brade, A.~Poddey, and W.~Branz, ``{Validation of
  automated driving -- A structured analysis and survey of approaches},'' in
  \emph{{Proc. Workshop Fahrerassistenzsysteme}}, {Uni-DAS e.V.}, Ed., 2020,
  pp. 64--73.

\bibitem{ISODIS21448_2021}
\emph{{Road Vehicles: Safety of the Intended Functionality}}, ISO Std. 21\,448,
  2021.

\bibitem{Schuldt_2017}
F.~Schuldt, ``{Ein Beitrag f{\"u}r den methodischen Test von automatisierten
  Fahrfunktionen mit Hilfe von virtuellen Umgebungen},'' Ph.D. dissertation,
  {Techn. Univ. Braunschweig}, Braunschweig, Germany, 2017.

\bibitem{Wood_2019}
\BIBentryALTinterwordspacing
M.~{Wood \textit{et~al.}}, ``{Safety first for automated driving},'' 2019.
  Accessed: Nov. 12, 2021. [Online]. Available:
  \url{https://www.daimler.com/dokumente/innovation/sonstiges/safety-first-for-automated-driving.pdf}
\BIBentrySTDinterwordspacing

\bibitem{Stolte_2015}
T.~Stolte, A.~Reschka, G.~Bagschik, and M.~Maurer, ``{Towards automated
  driving: Unmanned protective vehicle for highway hard shoulder road works},''
  in \emph{{Proc. IEEE 18th Int. Conf. Intell. Transp. Syst.}}, 2015, pp.
  672--677, \href{https://doi.org/10.1109/ITSC.2015.115}{doi:
  10.1109/ITSC.2015.115}.

\bibitem{ENABLE_S3_Method_2019}
\BIBentryALTinterwordspacing
{ENABLE-S3 Project Consortium}, ``{Testing and validation of highly automated
  systems: Summary of results},'' 2019, Accessed: Nov. 12, 2021. [Online].
  Available:
  \url{https://drive.google.com/file/d/15c1Oe69dpvW5dma8-uS8hev17x-6V3zU/view}
\BIBentrySTDinterwordspacing

\bibitem{PEGASUSMethod_2019}
\BIBentryALTinterwordspacing
{PEGASUS Project Consortium}, ``{PEGASUS method: An overview},'' 2019,
  Accessed: Nov. 12, 2021. [Online]. Available:
  \url{https://www.pegasusprojekt.de/files/tmpl/Pegasus-Abschlussveranstaltung/PEGASUS-Gesamtmethode.pdf}
\BIBentrySTDinterwordspacing

\bibitem{SL45_Homepage}
\BIBentryALTinterwordspacing
{SET Level Project Consortium}, ``{SET Level -- Simulation-based development
  and testing of automated driving},'' Accessed: Nov. 12, 2021. [Online].
  Available: \url{https://setlevel.de/en}
\BIBentrySTDinterwordspacing

\bibitem{VVM_Homepage}
\BIBentryALTinterwordspacing
{VVM Project Consortium}, ``{VVM -- Verification and validation methods for
  automated vehicles in urban environments},'' Accessed: Nov. 12, 2021.
  [Online]. Available: \url{https://www.vvm-projekt.de/en/}
\BIBentrySTDinterwordspacing

\bibitem{Menzel_2018a}
T.~Menzel, G.~Bagschik, and M.~Maurer, ``{Scenarios for development, test and
  validation of automated vehicles},'' in \emph{{Proc. IEEE Intell. Vehicles
  Symp.}}, 2018, pp. 1821--1827,
  \href{https://doi.org/10.1109/IVS.2018.8500406}{doi:
  10.1109/IVS.2018.8500406}.

\bibitem{ISO26262_2018}
\emph{{Road Vehicles -- Functional Safety}}, ISO Std. 26\,262, 2018.

\bibitem{Menzel_2019}
T.~Menzel, G.~Bagschik, L.~Isensee, A.~Schomburg, and M.~Maurer, ``{From
  functional to logical scenarios: Detailing a keyword-based scenario
  description for execution in a simulation environment},'' in \emph{{Proc.
  IEEE Intell. Vehicles Symp.}}, 2019, pp. 2383--2390,
  \href{https://doi.org/10.1109/IVS.2019.8814099}{doi:
  10.1109/IVS.2019.8814099}.

\bibitem{Gelder_2017}
E.~de~Gelder and J.-P. Paardekooper, ``{Assessment of Automated Driving Systems
  using real-life scenarios},'' in \emph{{Proc. IEEE Intell. Vehicles Symp.}},
  2017, pp. 589--594, \href{https://doi.org/10.1109/IVS.2017.7995782}{doi:
  10.1109/IVS.2017.7995782}.

\bibitem{Putz_2017}
A.~P{\"u}tz, A.~Zlocki, J.~Bock, and L.~Eckstein, ``{System validation of
  highly automated vehicles with a database of relevant traffic scenarios},''
  in \emph{{12th ITS Eur. Congr.}}, 2017.

\bibitem{waymo_SafetyReport_2020}
\BIBentryALTinterwordspacing
{Waymo LLC}, ``{Waymo Safety Report},'' Mountain View, CA, USA, 2020, Accessed:
  Nov. 12, 2021. [Online]. Available:
  \url{https://waymo.com/safety/safety-report}
\BIBentrySTDinterwordspacing

\bibitem{Weber_2020}
N.~Weber, D.~Frerichs, and U.~Eberle, ``{A simulation-based, statistical
  approach for the derivation of concrete scenarios for the release of highly
  automated driving functions},'' in \emph{{AmE 2020 -- Automot. meets
  Electron.; 11th GMM-Symp.}}\hskip 1em plus 0.5em minus 0.4em\relax {VDE
  Verlag}, 2020, pp. 116--121.

\bibitem{Bagschik_2018b}
G.~Bagschik, T.~Menzel, and M.~Maurer, ``{Ontology based scene creation for the
  development of automated vehicles},'' in \emph{{Proc. IEEE Intell. Vehicles
  Symp.}}, 2018, pp. 1813--1820,
  \href{https://doi.org/10.1109/IVS.2018.8500632}{doi:
  10.1109/IVS.2018.8500632}.

\bibitem{Bagschik_2018}
G.~Bagschik, T.~Menzel, C.~K{\"o}rner, and M.~Maurer, ``{Wissensbasierte
  Szenariengenerierung f{\"u}r Betriebsszenarien auf deutschen Autobahnen},''
  in \emph{{Proc. Workshop Fahrerassistenzsysteme}}, {Uni-DAS e.V.}, Ed., 2018,
  pp. 1--14.

\bibitem{Neurohr_2020}
C.~Neurohr, L.~Westhofen, T.~Henning, T.~de~Graaff, E.~M{\"o}hlmann, and
  E.~B{\"o}de, ``{Fundamental considerations around scenario-based testing for
  automated driving},'' in \emph{{Proc. IEEE Intell. Vehicles Symp.}}, 2020,
  pp. 121--127, \href{https://doi.org/10.1109/IV47402.2020.9304823}{doi:
  10.1109/IV47402.2020.9304823}.

\bibitem{Steimle_2021}
M.~Steimle, T.~Menzel, and M.~Maurer, ``{Toward a consistent taxonomy for
  scenario-based development and test approaches for automated vehicles: A
  proposal for a structuring framework, a basic vocabulary, and its
  application},'' \emph{{IEEE Access}}, vol.~9, pp. 147\,828--147\,854, 2021,
  \href{https://doi.org/10.1109/ACCESS.2021.3123504}{doi:
  10.1109/ACCESS.2021.3123504}.

\bibitem{Baumann_2021}
D.~Baumann, R.~Pfeffer, and E.~Sax, ``{Automatic generation of critical test
  cases for the development of highly automated driving functions},'' in
  \emph{{Proc. IEEE 93rd Veh. Technol. Conf.}}, 2021, pp. 1--5,
  \href{https://doi.org/10.1109/VTC2021-Spring51267.2021.9448686}{doi:
  10.1109/VTC2021-Spring51267.2021.9448686}.

\bibitem{Amersbach_2020}
C.~T. Amersbach, ``{Functional Decomposition Approach - Reducing the Safety
  Validation Effort for Highly Automated Driving},'' Ph.D. dissertation,
  {Techn. Univ. Darmstadt}, Darmstadt, Germany, 2020.

\bibitem{Balci_2010}
O.~Balci, ``{Golden rules of verification, validation, testing, and
  certification of modeling and simulation applications},'' \emph{{SCS M{\&}S
  Mag.}}, vol.~4, 2010.

\bibitem{Law_2019}
A.~M. Law, ``{How to build valid and credible simulation models},'' in
  \emph{{Proc. Winter Simul. Conf.}}, 2019, pp. 1402--1414,
  \href{https://doi.org/10.1109/WSC40007.2019.9004789}{doi:
  10.1109/WSC40007.2019.9004789}.

\bibitem{Walden_2015}
D.~D. Walden, G.~J. Roedler, K.~J. Forsberg, R.~D. Hamelin, and T.~M. Shortell,
  Eds., \emph{{Systems engineering handbook: A guide for system life cycle
  processes and activities}}, 4th~ed.\hskip 1em plus 0.5em minus 0.4em\relax
  Hoboken, NJ: Wiley, 2015.

\bibitem{NASA_7009A_2016}
\emph{{Standard for Models and Simulations}}, {National Aeronautics and Space
  Administration} Std. 7009A, 2016.

\bibitem{Klemmer_2011}
J.~Klemmer, J.~Lauer, V.~Formanski, R.~Fontaine, P.~Kilian, S.~Sinsel,
  A.~Erbes, and J.~Z{\"a}pf, ``{Definition and application of a standard
  verification and validation process for dynamic vehicle simulation models},''
  \emph{{SAE Int. J. Mater. Manuf.}}, vol.~4, no.~1, pp. 743--758, 2011,
  \href{https://doi.org/10.4271/2011-01-0519}{doi: 10.4271/2011-01-0519}.

\bibitem{Viehof_2018b}
M.~Viehof and H.~Winner, ``{Research methodology for a new validation concept
  in vehicle dynamics},'' \emph{{Automot. and Engine Technol.}}, vol.~3, no.
  1-2, pp. 21--27, 2018, \href{https://doi.org/10.1007/s41104-018-0024-1}{doi:
  10.1007/s41104-018-0024-1}.

\bibitem{Schuldt_2015}
F.~Schuldt, T.~Menzel, and M.~Maurer, ``{Eine Methode f{\"u}r die Zuordnung von
  Testf{\"a}llen f{\"u}r automatisierte Fahrfunktionen auf X-in-the-Loop
  Verfahren im modularen virtuellen Testbaukasten},'' in \emph{{Proc. Workshop
  Fahrerassistenzsysteme}}, {Uni-DAS e.V.}, Ed., 2015, pp. 171--182.

\bibitem{Bode_2018}
E.~B{\"o}de, M.~B{\"u}ker, U.~Eberle, M.~Fr{\"a}nzle, S.~Gerwinn, and
  B.~Kramer, ``{Efficient splitting of test and simulation cases for the
  verification of highly automated driving functions},'' in \emph{{Int. Conf.
  Comput. Saf. 2018}}, 2018, pp. 139--153.

\bibitem{Antkiewicz_2020}
M.~Antkiewicz, M.~Kahn, M.~Ala, and K.~{Czarnecki \textit{et al.}}, ``{Modes of
  Automated Driving System Scenario Testing: Experience Report and
  Recommendations},'' \emph{{SAE Int. J. Advances {\&} Curr. Prac. in Mobility
  2(4):2248-2266}}, 2020, \href{https://doi.org/10.4271/2020-01-1204}{doi:
  10.4271/2020-01-1204}.

\bibitem{Strasser_2012}
B.~Strasser, ``{Vernetzung von Test- und Simulationsmethoden f{\"u}r die
  Entwicklung von Fahrerassistenzsystemen},'' Ph.D. dissertation, {Techn. Univ.
  M{\"u}nchen}, Munich, Germany, 2012.

\bibitem{NeumannCosel_2014}
K.~von Neumann-Cosel, ``{Virtual Test Drive: Simulation umfeldbasierter
  Fahrzeugfunktionen},'' Ph.D. dissertation, {Techn. Univ. M{\"u}nchen},
  Munich, Germany, 2014.

\bibitem{Stellet_2015}
J.~E. Stellet, M.~R. Zofka, J.~Schumacher, T.~Schamm, F.~Niewels, and J.~M.
  Z{\"o}llner, ``{Testing of advanced driver assistance towards automated
  driving: A survey and taxonomy on existing approaches and open questions},''
  in \emph{{Proc. IEEE 18th Int. Conf. Intell. Transp. Syst.}}, 2015, pp.
  1455--1462, \href{https://doi.org/10.1109/ITSC.2015.236}{doi:
  10.1109/ITSC.2015.236}.

\bibitem{Steimle_2019}
\BIBentryALTinterwordspacing
M.~Steimle, T.~Menzel, and M.~Maurer, ``{A method for classifying test bench
  configurations in a scenario-based test approach for automated vehicles},''
  2019, Accessed: Nov. 12, 2021. [Online]. Available:
  \url{https://arxiv.org/abs/1905.09018v1}
\BIBentrySTDinterwordspacing

\bibitem{Steimle_2021a}
\BIBentryALTinterwordspacing
M.~Steimle, N.~Weber, and M.~Maurer, ``{Toward generating sufficiently valid
  test case results: A method for systematically assigning test cases to test
  bench configurations in a scenario-based test approach for automated
  vehicles},'' 2021, Accessed: Nov. 12, 2021. [Online]. Available:
  \url{https://arxiv.org/abs/2109.03146v2}
\BIBentrySTDinterwordspacing

\bibitem{Stellet_2021}
J.~E. Stellet, private communication, Oct. 2021.

\bibitem{Bubb_1993}
H.~Bubb and H.~Schmidtke, ``{Systemstruktur},'' in \emph{{Ergonomie}},
  H.~Schmidtke, Ed.\hskip 1em plus 0.5em minus 0.4em\relax {Carl
  Hanser-Verlag}, 1993, pp. 305--332.

\bibitem{Schaermann_2017}
A.~Schaermann, A.~Rauch, N.~Hirsenkorn, T.~Hanke, R.~Rasshofer, and E.~Biebl,
  ``{Validation of vehicle environment sensor models},'' in \emph{{Proc. IEEE
  Intell. Vehicles Symp.}}, 2017, pp. 405--411,
  \href{https://doi.org/10.1109/IVS.2017.7995752}{doi:
  10.1109/IVS.2017.7995752}.

\bibitem{Holder_2018}
M.~Holder, P.~Rosenberger, H.~Winner, T.~Dhondt, V.~P. Makkapati, M.~Maier,
  H.~Schreiber, Z.~Magosi, Z.~Slavik, O.~Bringmann, and W.~Rosenstiel,
  ``{Measurements revealing challenges in radar sensor modeling for virtual
  validation of autonomous driving},'' in \emph{{Proc. IEEE 21st Int. Conf.
  Intell. Transp. Syst.}}, 2018, pp. 2616--2622,
  \href{https://doi.org/10.1109/ITSC.2018.8569423}{doi:
  10.1109/ITSC.2018.8569423}.

\bibitem{Bock_2018}
J.~Bock, R.~Krajewski, L.~Eckstein, J.~Klimke, J.~Sauerbier, and A.~Zlocki,
  ``{Data basis for scenario-based validation of HAD on highways},'' in
  \emph{{27th Aachen Colloq. Automobile 2018}}, 2018.

\bibitem{Weber_2019}
H.~Weber, J.~Bock, J.~Klimke, {Roesener Christian}, J.~Hiller, R.~Krajewski,
  {Zlocki Adrian}, L.~Eckstein, and Krajew, ``{A framework for definition of
  logical scenarios for safety assurance of automated driving},''
  \emph{{Traffic Injury Prevention}}, vol. 20:sup1, pp. 65--70, 2019,
  \href{https://doi.org/10.1080/15389588.2019.1630827}{doi:
  10.1080/15389588.2019.1630827}.

\bibitem{Scholtes_2021}
M.~Scholtes, L.~Westhofen, L.~R. Turner, K.~Lotto, M.~Schuldes, H.~Weber,
  N.~Wagener, C.~Neurohr, M.~Bollmann, F.~K{\"o}rtke, J.~Hiller, M.~Hoss,
  J.~Bock, and L.~Eckstein, ``6-layer model for a structured description and
  categorization of urban traffic and environment,'' \emph{{IEEE Access}},
  vol.~9, pp. 59\,131--59\,147, 2021,
  \href{https://doi.org/10.1109/ACCESS.2021.3072739}{doi:
  10.1109/ACCESS.2021.3072739}.

\bibitem{Matthaei_2015}
R.~Matthaei and M.~Maurer, ``{Autonomous driving -- A top-down-approach},'' in
  \emph{{at - Automatisierungstechnik}}, G.~Bretthauer, Ed., vol. 63 - Issue 3,
  2015, pp. 155--167, \href{https://doi.org/10.1515/auto-2014-1136}{doi:
  10.1515/auto-2014-1136}.

\bibitem{Matthaei_2015b}
R.~Matthaei, ``{Wahrnehmungsgest{\"u}tzte Lokalisierung in fahrstreifengenauen
  Karten f{\"u}r Assistenzsysteme und automatisches Fahren in urbaner
  Umgebung},'' Ph.D. dissertation, {Techn. Univ. Braunschweig}, Braunschweig,
  Germany, 2015.

\bibitem{Ulbrich_2017}
\BIBentryALTinterwordspacing
S.~Ulbrich, A.~Reschka, J.~Rieken, S.~Ernst, G.~Bagschik, F.~Dierkes, M.~Nolte,
  and M.~Maurer, ``{Towards a functional system architecture for automated
  vehicles},'' 2017, Accessed: Nov. 12, 2021. [Online]. Available:
  \url{https://arxiv.org/abs/1703.08557v2}
\BIBentrySTDinterwordspacing

\bibitem{Ulbrich_2015b}
S.~Ulbrich, T.~Menzel, A.~Reschka, F.~Schuldt, and M.~Maurer, ``{Defining and
  substantiating the terms scene, situation, and scenario for automated
  driving},'' in \emph{{Proc. IEEE 18th Int. Conf. Intell. Transp. Syst.}},
  2015, pp. 982--988, \href{https://doi.org/10.1109/ITSC.2015.164}{doi:
  10.1109/ITSC.2015.164}.

\bibitem{Berg_2014}
G.~Berg, ``{Das Vehicle in the Loop - Ein Werkzeug f{\"u}r die Entwicklung und
  Evaluation von sicherheitskritischen Fahrerassistenzsystemen},'' Ph.D.
  dissertation, {Univ. der Bundeswehr M{\"u}nchen}, Munich, Germany, 2014.

\bibitem{Hallerbach_2018}
S.~Hallerbach, Y.~Xia, U.~Eberle, and F.~Koester, ``{Simulation-based
  identification of critical scenarios for cooperative and automated
  vehicles},'' \emph{{SAE Int. J. CAV}}, vol.~1, no.~2, pp. 93--106, 2018,
  \href{https://doi.org/10.4271/2018-01-1066}{doi: 10.4271/2018-01-1066}.

\bibitem{ASAMOSI_2021}
\BIBentryALTinterwordspacing
{ASAM e.V.}, ``{ASAM OSI},'' Accessed: Nov. 12, 2021. [Online]. Available:
  \url{https://www.asam.net/standards/detail/osi/}
\BIBentrySTDinterwordspacing

\bibitem{OSIDocumentation_2021}
\BIBentryALTinterwordspacing
------, ``{Open Simulation Interface's documentation},'' Accessed: Nov. 12,
  2021. [Online]. Available:
  \url{https://opensimulationinterface.github.io/osi-documentation/index.html}
\BIBentrySTDinterwordspacing

\bibitem{FMIHomepage_2021}
\BIBentryALTinterwordspacing
{Modelica Association c/o PELAB, IDA, Link{\"o}pings Universitet},
  ``{Functional Mock-up Interface},'' Accessed: Nov. 12, 2021. [Online].
  Available: \url{https://fmi-standard.org/}
\BIBentrySTDinterwordspacing

\bibitem{SSPHomepage_2021}
\BIBentryALTinterwordspacing
------, ``{System Structure and Parameterization},'' Accessed: Nov. 12, 2021.
  [Online]. Available: \url{https://ssp-standard.org/}
\BIBentrySTDinterwordspacing

\bibitem{Kramer_2021}
B.~Kramer, G.~Ehmen, and T.~Koopmann, private communication, Jan. 2021.

\bibitem{Frerichs_2021}
\BIBentryALTinterwordspacing
D.~Frerichs, A.~Gr{\"a}tz, and B.~Kramer, ``{It's all about trust -- Simulation
  credibility},'' 2021, Accessed: Nov. 12, 2021. [Online]. Available:
  \url{https://setlevel.de/assets/mid-term-presentation/Simulation-Credibility.pdf}
\BIBentrySTDinterwordspacing

\end{thebibliography}

\begin{IEEEbiography}[{\includegraphics[width=1in,height=1.25in,clip,keepaspectratio]{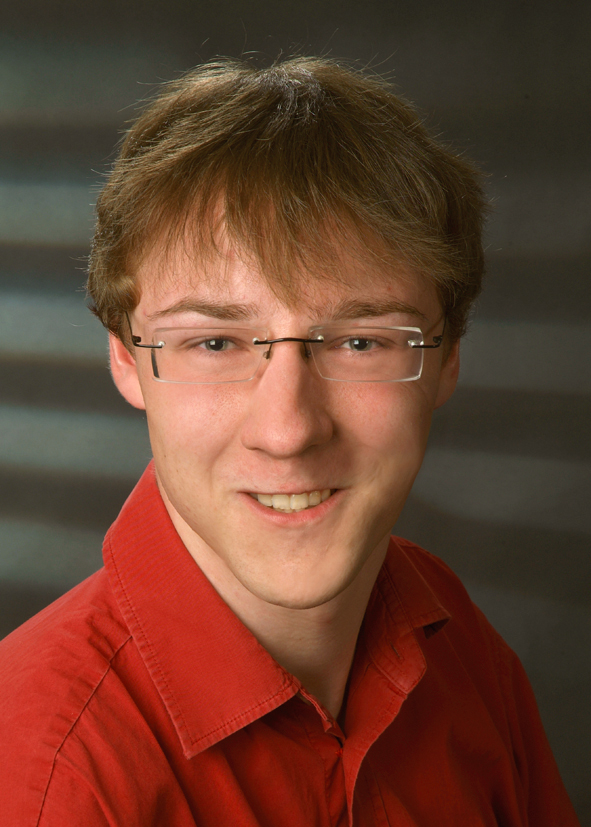}}]{Markus Steimle} received the B.Eng. degree in electrical engineering and information technology from the Landshut University of Applied Sciences, Landshut, Germany, in 2014, and the M.Sc. degree in automotive software engineering from the Technical University of Munich, Munich, Germany, in 2016. 
He is currently pursuing the Ph.D. degree with the Institute of Control Engineering, Technische Universität Braunschweig, Braunschweig, Germany. 
His main research interests include scenario-based verification and validation of automated vehicles, focusing on the use of simulative testing methods.
\end{IEEEbiography}

\begin{IEEEbiography}[{\includegraphics[width=1in,height=1.25in,clip,keepaspectratio]{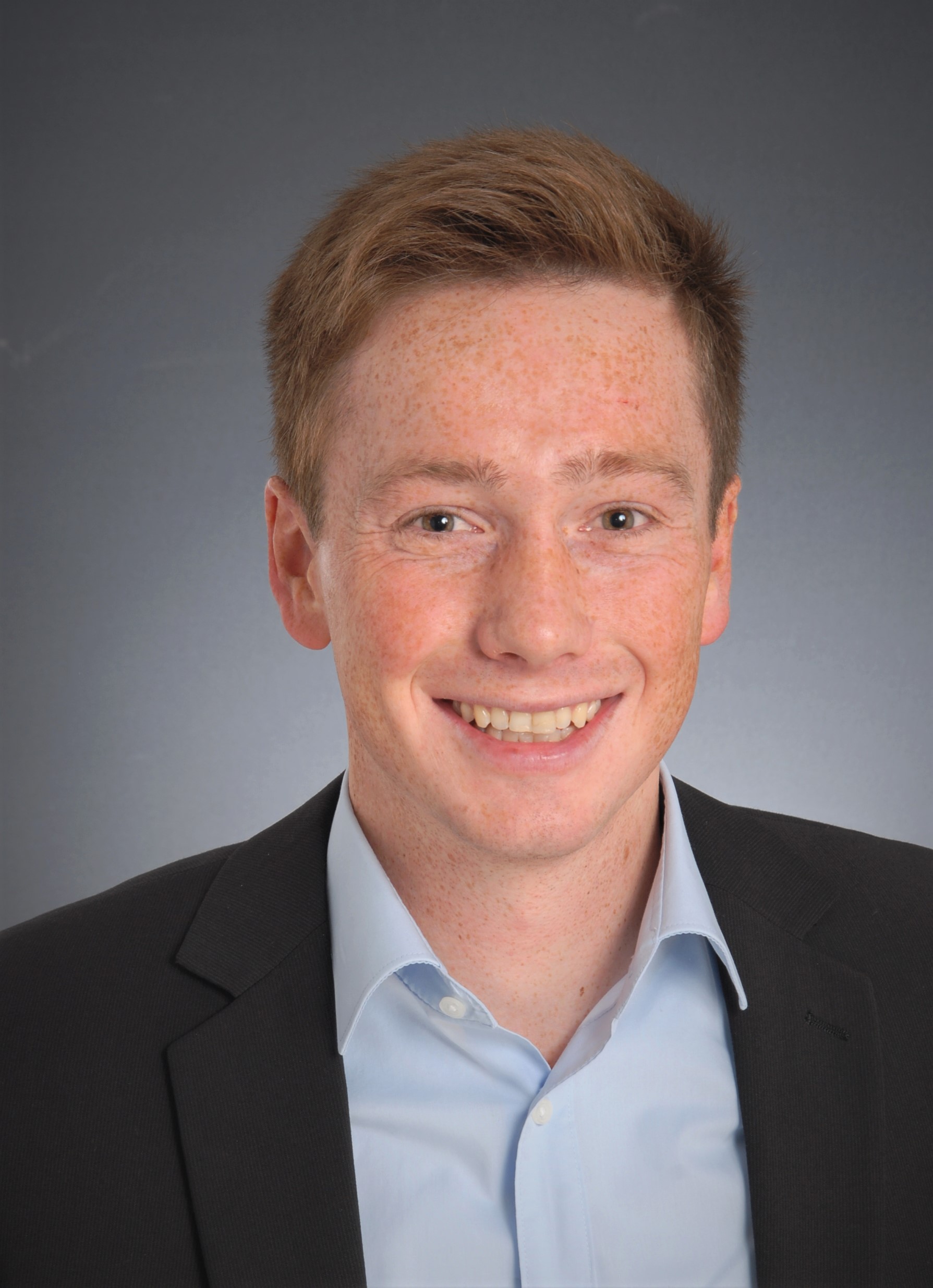}}]{Nico Weber} received the B.Sc. and the M.Sc. degree in mechanical and process engineering from the Technical University of Darmstadt, Darmstadt, Germany, in 2017 and 2019, respectively.
	Since 2019, he has been an external Ph.D. student at the Control Systems and Mechatronics Laboratory (rtm) of the Technical University of Darmstadt, Germany, conducting his research at Opel Automobile, Stellantis. 
	His main research interests include scenario-based verification and validation of automated vehicles, focusing on the use of simulative development and testing methods.
\end{IEEEbiography}

\begin{IEEEbiography}[{\includegraphics[width=1in,height=1.25in,clip,keepaspectratio]{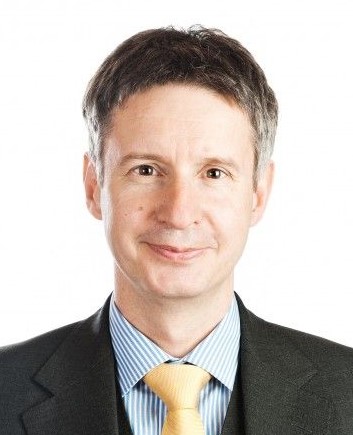}}]{Markus Maurer} received the Diploma degree in electrical engineering from the Technische Universität München, in 1993, and the Ph.D. degree in automated driving from the Group of Prof. E. D. Dickmanns, Universität der Bundeswehr München, in 2000. 
From 1999 to 2007, he was a Project Manager and the Head of the Development Department of Driver Assistance Systems, Audi. 
Since 2007, he has been a Full Professor of automotive electronics systems with the Institute of Control Engineering, Technische Universität Braunschweig.
His research interests include both functional and systemic aspects of automated road vehicles.
(Picture: Jessen Oestergaard / Daimler and Benz Foundation)
\end{IEEEbiography}

\EOD

\end{document}